\documentclass[floats,floatfix,showpacs,amssymb,prd,superscriptaddress,nofootinbib,aps,notitlepage,reprint]{revtex4-1}

\usepackage{graphicx, epsfig, amssymb} 
\usepackage{amsmath, amsfonts}
\usepackage{bm} 

\usepackage[linktocpage]{hyperref}
\usepackage[caption=false]{subfig}
\usepackage[usenames]{color}

\def\be{\begin{equation}}
\def\ee{\end{equation}}

\newcommand{\bea}{\begin{eqnarray}}
\newcommand{\eea}{\end{eqnarray}}
\newcommand{\ben}{\begin{enumerate}}
\newcommand{\een}{\end{enumerate}}
\newcommand{\bi}{\begin{itemize}}
\newcommand{\ei}{\end{itemize}}

\newcommand{\nn}{\nonumber}

\def\ga{\mathrel{\raise.3ex\hbox{$>$\kern-.75em\lower1ex\hbox{$\sim$}}}}
\def\la{\mathrel{\raise.3ex\hbox{$<$\kern-.75em\lower1ex\hbox{$\sim$}}}}

\def\l{\left}
\def\r{\right}
\def\be{\begin{equation}}
\def\ee{\end{equation}}

\def\I_M{{I_{\scriptscriptstyle M\times M}}}

\def\be{\begin{equation}}
\def\ee{\end{equation}}
\def\bea{\begin{eqnarray}}
\def\eea{\end{eqnarray}}
\newcommand{\beq}{\begin{eqnarray}}
\newcommand{\eeq}{\end{eqnarray}}
\def\pa{\partial}

\begin{document}\title{\large Astrophysical signatures of boson stars: Quasinormal modes and inspiral resonances}

\author{Caio F. B. Macedo}\email{caiomacedo@ufpa.br}
\affiliation{Faculdade de F\'{\i}sica, Universidade 
Federal do Par\'a, 66075-110, Bel\'em, Par\'a, Brazil.}
\affiliation{CENTRA, Departamento de F\'{\i}sica, 
Instituto Superior T\'ecnico, Universidade T\'ecnica de Lisboa - UTL,
Avenida Rovisco Pais 1, 1049 Lisboa, Portugal.}

\author{Paolo Pani}\email{paolo.pani@ist.utl.pt}
\affiliation{CENTRA, Departamento de F\'{\i}sica, 
Instituto Superior T\'ecnico, Universidade T\'ecnica de Lisboa - UTL,
Avenida Rovisco Pais 1, 1049 Lisboa, Portugal.}
\affiliation{Institute for Theory and Computation, Harvard-Smithsonian
CfA, 60 Garden Street, Cambridge, Massachusetts 02138, USA}

\author{Vitor Cardoso}\email{vitor.cardoso@ist.utl.pt}
\affiliation{CENTRA, Departamento de F\'{\i}sica, 
Instituto Superior T\'ecnico, Universidade T\'ecnica de Lisboa - UTL,
Avenida Rovisco Pais 1, 1049 Lisboa, Portugal.}
\affiliation{Perimeter Institute for Theoretical Physics
Waterloo, Ontario N2J 2W9, Canada.}
\affiliation{Faculdade de F\'{\i}sica, Universidade 
Federal do Par\'a, 66075-110, Bel\'em, Par\'a, Brazil.}
\affiliation{Department of Physics and Astronomy, The University of Mississippi, University, Mississippi 38677, USA.}

\author{Lu\'is C. B. Crispino}\email{crispino@ufpa.br}
\affiliation{Faculdade de F\'{\i}sica, Universidade 
Federal do Par\'a, 66075-110, Bel\'em, Par\'a, Brazil.}

\begin{abstract}
Compact bosonic field configurations, or boson stars, are promising dark matter candidates which have been invoked as
an alternative description for the supermassive compact objects, in active galactic nuclei. 
Boson stars can be comparable in size and mass to supermassive and they might be hard to distinguish by electromagnetic observations.
However, boson stars do not possess an event horizon and their global spacetime structure is different from that of a black hole. This leaves a characteristic imprint in the gravitational-wave emission, which can be used as a discriminant between black holes and other horizonless compact objects.
Here we perform a detailed study of boson stars and their gravitational-wave signatures in a fully relativistic setting, a study which was lacking
in the existing literature in many respects. We construct several fully relativistic boson star configurations, and we analyze their geodesic structure and free oscillation spectra, or quasinormal modes.
We explore the gravitational and scalar response of boson star spacetimes to an inspiraling stellar-mass object and compare it to its black hole counterpart. 
We find that a generic signature of compact boson stars is the resonant-mode excitation by a small compact object on stable quasicircular geodesic motion.  
\end{abstract}


\date{\today}

\maketitle


\section{Introduction}
We recently investigated gravitational-wave signatures of stellar-size objects orbiting around supermassive, dark matter configurations~\cite{Macedo:2013qea}. These extreme mass-ratio inspirals (EMRIs) were studied both inside and outside the supermassive object. The gravitational radiation output during the inner inspiral was treated by a Newtonian approach, which included accretion and gravitational drag, whereas the outer inspiral was described at the fully relativistic level. To model the motion of the particle outside the object, specific relativistic models have to be used and, in Ref.~\cite{Macedo:2013qea}, we considered boson star (BS) configurations, which are viable and promising dark matter candidates. Here we extend our study and discuss in detail some of the results briefly presented in Ref.~\cite{Macedo:2013qea}.

BSs are compact configurations satisfying the Einstein-Klein-Gordon equations, prevented from total collapse through the Heinsenberg uncertainty principle (for reviews on the subject see~\cite{Jetzer:1991jr,Schunck:2003kk,Liebling:2012fv}). They have been claimed in the literature as promising horizonless black hole (BH) mimickers, being possible star candidates for supermassive objects.
BSs can be classified~\cite{Schunck:2003kk} according to the scalar potential{, namely $V(\Phi)$ (see Sec. \ref{sec:eineq}),} in the Klein-Gordon Lagrangian. In this paper, we shall discuss some of the most popular BS models:
\begin{itemize}
\item {Mini boson stars}, for which the scalar potential is given by $V(\Phi)=\mu^2|\Phi |^2$, where $\mu$ is the scalar field mass. The maximum mass for this BS model is given by the so-called Kaup limit $M_{max}\approx 0.633 m_{\rm P}^2/\mu$, with $m_{\rm P}$ being the Planck mass~\cite{Kaup:1968zz,Ruffini:1969qy}. For typical values of $\mu$, this mass limit is much smaller than the Chandrasekhar limit for a fermion star, approximately $m_{\rm P}^3/\mu^2$. Nevertheless, despite their name, mini BSs may have a total mass compatible with that observed in active galactic nuclei~\cite{Schunck:2003kk}. This happens for ultralight boson masses $\mu$, as those motivated by string axiverse scenarios~\cite{Arvanitaki:2009fg}.
\item {Massive boson stars}, for which the scalar potential has an additional quartic scalar field term, $V(\Phi)=\mu^2|\Phi|^2+\lambda |\Phi|^4/2$~\cite{Colpi:1986ye}. Depending on the value of $\lambda$, the maximum mass can be comparable to the Chandrasekhar limit. For $\lambda \gg \mu^2/m_{\rm P}^2$ one can estimate $M_{max}\approx 0.062\lambda^{1/2}m_{\rm P}^3/\mu^2$. 
\item {Solitonic boson stars}, for which $V(\Phi)=\mu^2|\Phi|^2 (1-2|\Phi|^2/\sigma_0^2)^2$, where $\sigma_0$ is a constant~\cite{Friedberg:1986tq}. This potential supports confined nondispersive solutions with finite mass, even in the absence of gravity.
The total mass of the star depends on $\sigma_0$ and $M_{max}\approx 0.0198 m_{\rm P}^4/(\mu \sigma_0^2)$. This model (also known in the literature as nontopological solitonic stars) allows for supermassive objects with $M\sim 10^6 M_\odot$ even in the presence of heavy bosons with $\mu\sim\sigma_0\sim 500$~GeV. 
\end{itemize}
Other types of BSs can be obtained using different scalar self-potentials; see Ref.~\cite{Schunck:2003kk} for a more detailed list.

The emission spectra from a simple accretion disk model around BSs was studied in Refs. \cite{Torres:2002td,Guzman:2005bs}. It was shown that, depending on the BS model and on the compactness, spherically symmetric massive BSs can be indistinguishable from Schwarzschild BHs. In this sense, BS can supplant BHs as supermassive objects. Ways to discriminate BSs from BHs have been studied in the literature, such as the K$\alpha$ iron line profile from accretion disks \cite{Lu:2002vm} (see also \cite{Bambi:2013hza} for other compact objects) and gravitational lensing \cite{BinNun:2013fe} (see also Ref. \cite{Schunck:2003kk}).

Despite the vast existing literature on their dynamical features (cf. the recent review \cite{Liebling:2012fv}), a detailed study on the astrophysical signatures of BSs in a fully relativistic setting is missing. The scope of the present paper is to fill this gap. We study dynamical BSs in order to identify possible smoking guns of horizonless compact objects
and of compact dark matter configurations, extending previous studies in several directions. 

After giving the necessary formalism in Sec.~\ref{sec:equations}, we explore the three different types of BSs discussed above in Sec.~\ref{sec:background}. 
The spacetimes are obtained using the full Einstein equation, without any approximation scheme. Our results agree very well with the ones presented in the literature \cite{Ruffini:1969qy,Colpi:1986ye,Friedberg:1986tq}.

In Sec.~\ref{sec:geodesics} we characterize circular geodesics in BS spacetimes. 
In particular, even though a BS does not possess a well-defined surface and stable circular geodesics may exist even \emph{inside} the star, we find some upper bound on the angular frequency as measured by (static) asymptotic observers. 
In Sec.~\ref{sec:qnms} we compute the fundamental quasinormal modes (QNMs) of various BS models and show that there exists a class of low-frequency modes. In Sec.~\ref{sec:PP} we show that these modes can be excited by point particles in quasicircular geodesic motion. This is a striking difference from the BH case, where the QNMs can only be excited by particles plunging into the BH and not during the inspiral.

The results of Sec.~\ref{sec:qnms} are complementary to those of Refs.~\cite{Kojima:1991np,Yoshida:1994xi}, where the QNMs of mini BS configurations were computed using a Wentzel-Kramers-Brillouin (WKB) approximation (see also Ref.~\cite{LoraClavijo:2010xc} where the scalar QNMs of BS models in the probe limit were computed). We extend those results by considering several BS models and by computing the proper modes with more sophisticated methods that do not rely on any approximation scheme. More specifically, we focus on the quasibound state modes of the scalar field, and we argue that these are generic features of any BS configuration supported by a massive scalar field.

The results of Sec.~\ref{sec:PP} are complementary to--and in fact extend--the work by  Kesden \textit{et al.} \cite{Kesden:2004qx}, who calculated the approximated waveforms for gravitational waves emitted by particle inspirals from the Schwarzschild exterior to the interior of a nontopological soliton star.
As in Ref.~\cite{Kesden:2004qx}, here we have the broad goal of studying gravitational-wave emission by EMRIs around generic horizonless objects.
EMRIs are unique probes of the strong-curvature regime of general relativity (GR) and are also perfect test beds to put constraints on modified theories of gravity (see, e.g., Refs.~\cite{Yunes:2011aa,Pani:2011xj}). In addition to computing the gravitational and scalar energy fluxes in a consistent and fully relativistic approach for several BS models, we find that the absence of the ``one-way membrane'' (event horizon) opens up the possibility that the free oscillation modes of a BS
are measurably different from those of a BH, and they can even be resonantly excited by orbiting point particles.
Indeed, we find that orbiting stellar-mass objects around BSs generically excite a multitude of resonant frequencies,
{\it and} give rise to a signal which in its last stages bears no resemblance to chirp or ringdown signals typical of inspirals into BHs.
We have discussed the detectability of these resonances in Ref.~\cite{Macedo:2013qea}.

Our results might be interesting at various levels but, from a phenomenological standpoint, the main message is that gravitational waves do allow a discrimination between compact objects, in particular between BHs and BSs. 
We use the signature $(-,+,+,+)$ for the metric and natural units $\hbar=c=G=1$.

\section{Einstein's equation for a particle orbiting a boson star \label{sec:equations}}
\label{sec:eineq}
BSs are equilibrium self-gravitating solutions of the Einstein-Klein-Gordon theory:
\be
S=\int d^4 x \sqrt{-g} \left[\frac{R}{2\kappa} -g^{ab}\partial_a\Phi^*\partial_b\Phi-V(|\Phi|^2)\right] +S_{\rm matter},\nn
\ee
where $\kappa=8 \pi$ and $S_{\rm matter}$ denotes the action of any other matter field. From the action above, Einstein's equations read
\begin{equation}
R_{ab}-\frac{1}{2}g_{ab}R = \kappa\l( T_{ab}^\Phi + T_{ab}^{\rm matter} \r)\,,\label{eineq}\\
\end{equation}
where
\be
T^\Phi_{ab}=\partial_a\Phi^*\partial_b\Phi+\partial_b\Phi^*\partial_a\Phi-g_{ab}\l(\partial^c\Phi^*\partial_c\Phi+V(|\Phi|^2)\r)\,,
\ee
is the energy-momentum of the scalar field. The Klein-Gordon equation reads
\begin{equation}
 \frac{1}{\sqrt{-g}}\pa_a \l(\sqrt{-g}g^{ab}\pa_b\Phi\r)=\frac{d V}{d|\Phi|^2}\Phi\,,\label{eq:phieq}
\end{equation}
together with its complex conjugate.
%
\subsection{Background solutions}
\label{sec:bkgsol}
We will focus exclusively on spherically symmetric BSs and consider the background line element
\be
ds_0^2=-e^{v(r)}dt^2+e^{u(r)}dr^2+r^2 (d\theta^2+\sin^2\theta d\varphi^2)\,. \label{metric0}
\ee
The ansatz for the background scalar field reads~\cite{Liebling:2012fv}
\be
\Phi_0(t,r)\equiv \phi_0(r)e^{-i\omega t}\,,
\ee
where $\phi_0(r)$ is a real function. Although the scalar field is time dependent, the Einstein-Klein-Gordon system admits static and spherically symmetric metrics~\cite{Kaup:1968zz,Ruffini:1969qy,Colpi:1986ye,Gleiser:1988rq,Gleiser:1988ih,Schunck:2003kk}. With the ansatz above, the background field equations, obtained from \eqref{eineq}--\eqref{eq:phieq}, read
\bea
\frac{1}{r^2}\l(r\,e^{-u}\r)' -\frac{1}{r^2}&=&-\kappa\rho\,,\label{bg11}\\
e^{-u}\l(\frac{v'}{r}+\frac{1}{r^2}\r)-\frac{1}{r^2}&=&\kappa p_{\text{rad}}\,,\label{bg22}\\
\phi_0''+\l(\frac{2}{r}+\frac{v'-u'}{2}\r)\phi_0'&=&e^u\l({U_0}-\omega^2e^{-v}\r)\phi_0\,,\label{bgscalar}
\eea
where a prime denotes the derivative with respect to $r$, $U_0=U(\phi_0)$ and $U(\Phi)={d V}/{d|\Phi|^2}$. In the equations above, the density $\rho$, the radial pressure $p_{\text{rad}}$, and the tangential pressure $p_{\text{tan}}$ are given in terms of the stress-energy tensor of the scalar field, ${T^\Phi}_{ab}$. More specifically, 
\bea
\rho&\equiv&-{{T^\Phi}_t}^t=\omega^2e^{-v}\phi_0^2+e^{-u}(\phi_0')^2+V_0\,,\\
p_{\text{rad}}&\equiv&{{T^\Phi}_r}^r=\omega^2e^{-v}\phi_0^2+e^{-u}(\phi_0')^2-V_0\,,\\
p_{\text{tan}}&\equiv&{{T^\Phi}_\theta}^\theta=\omega^2e^{-v}\phi_0^2-e^{-u}(\phi_0')^2- V_0\,.
\eea
where $V_0=V(\phi_0)$. Unlike the case of perfect fluid stars, the complex scalar field behaves like an anisotropic fluid, $p_{\text{rad}}\neq p_{\text{tan}}$. Equations~\eqref{bg11}--\eqref{bgscalar} can be solved numerically with suitable boundary conditions (see Sec.~\ref{sec:background}) to obtain the background metric and scalar field configuration.
\subsection{Perturbations}
We are interested in the free oscillation spectrum of a BS as well as in the scalar field and metric perturbations induced by test particles on geodesic motion in the spherically symmetric spacetime described above. At first-order in the perturbations, the metric reads
\be
g_{ab}=g_{ab}^{(0)}+h_{ab}\,,
\ee
where $g_{ab}^{(0)}$ is given in Eq.~\eqref{metric0}. In the Regge-Wheeler gauge~\cite{Regge:1957td}, using a Fourier expansion, the first order perturbation $h_{ab}$ separates into the axial sector
\bea
h_{ab}^{axial}&=&\sum_{l\geq|m| }\int d\sigma\l(
\begin{array}{cccc}
0 & 0 & -\frac{1}{\sin\theta}h_0(r)\pa_\varphi & \sin\theta~ h_0(r)\pa_\theta\\
\star & 0 & -\frac{1}{\sin\theta}h_1(r)\pa_\varphi & \sin\theta~ h_1(r)\pa_\theta\\
\star & \star & 0 & 0\\
\star & \star & \star & 0
\end{array}\r)\nn\\
&\times & Y^{lm} e^{-i\sigma t}
\eea
and polar sector
\bea
&&h_{ab}^{polar}=\nn\\
&&\sum_{l\geq|m|}\int d\sigma\l(
\begin{array}{cccc}
e^v  H_0(r) &  i \sigma H_1(r)  & 0 & 0 \\
\star & e^u H_2(r) & 0 & 0\\
\star & \star & r^2 K(r) & 0\\
\star & \star & \star & r^2 \sin^2\theta K(r)
\end{array}\r)\nn\\
&\times& Y^{lm}e^{-i\sigma t}\,,
\eea
where $Y^{lm}\equiv Y^{lm}(\theta,\varphi)$ are the usual scalar spherical harmonics. Each metric and scalar field perturbation, e.g. $h_0(r)$, explicitly depends on the frequency $\sigma$ and on the wave numbers $l$ and $m$. The $\star$ symbol indicates the symmetric components, such that $h_{ab}=h_{ba}$.

At first order, the scalar field reads $\Phi =\Phi_0+\delta \Phi$, where $\Phi_0$ is the background scalar field defined above and 
\bea
\delta \Phi&=& \sum_{l\geq|m|}\int d\sigma  \frac{\phi_+(r)}{r}~Y^{lm}e^{-i(\sigma +\omega) t }\,,\label{pscal1}\\
\delta \Phi^*&=& \sum_{l\geq|m|}\int d\sigma\frac{\phi_-(r)}{r}~Y^{lm}e^{-i(\sigma -\omega) t}\,.\label{pscal2}
\eea
Note that the ansatz above differs from that used in Refs.~\cite{Kojima:1991np,Yoshida:1994xi}.  
The scalar field potential can be written as
\be
V= V_0+\sum_{l\geq|m|}\int d\sigma\,\delta V(r)~Y^{lm}e^{-i\sigma t}\,.
\ee
Likewise, for the first derivative
\be
\frac{d V}{d|\Phi|^2}=U_0(r)+\sum_{l\geq|m|}\int d\sigma\,\delta {{U}}(r)~ Y^{lm}e^{-i\sigma t}.
\ee
In the presence of matter fields other than the complex scalar, $T_{ab}^{\rm matter}$ also has to be expanded in tensorial harmonics~\cite{Zerilli:1971wd,Sago:2002fe}.
In the time domain, the matter stress-energy tensor of a particle in the $\theta=\pi/2$ plane reads
\bea
T^{\rm matter}_{ab}&=&\mu_p\frac{\dot{x}_{a}(t)\dot{x}_{b}(t)}{{r_p(t)}^2\dot{x}^t(t) }e^{-\frac{1}{2}(v+u)}\nn\\
&&\times\delta(r-r_p(t))\delta(\cos\theta)\delta(\varphi-\varphi_p(t)),\nn
\eea
where $\dot{x}^a\equiv(\dot{t}_p,\dot{r}_p,0,\dot{\phi}_p)$ and $\mu_p$ are the particle's four-velocity and mass, respectively.
\subsubsection{Axial s	ector}\label{axials}
As discussed in Ref.~\cite{Kojima:1991np}, perturbations of the scalar field have even parity, so they couple only with polar gravitational perturbations.
Thus, gravitational axial perturbations decouple and they are described by the linearized Einstein equations \eqref{eineq}, namely,
\bea
e^{-u}h_1'+i\sigma e^{-v}h_0+\frac{1}{2}\kappa(p_{\text{rad}}-\rho)h_1+\frac{2}{r^2} m(r) h_1&=&P_{\sigma l m} \,,\nn\\ \label{thetaphicom}\\
-i\sigma h_0'+\frac{2i\sigma}{r}h_0 - \l[\sigma^2-\frac{e^v}{r^2}\l(l(l+1)-2\r)\r]h_1&=&P^r_{\sigma l m} \,,\nn\\   \label{rphicom}\\
i\sigma h_1' +h_0''-\frac{1}{2}\kappa r e^u(\rho+p_{\text{rad}})(h_0'+ i \sigma h_1)+\frac{2i\sigma}{r}h_1&&\nn\\
+h_0e^u \l[\kappa(p_{\text{rad}}+\rho)-\frac{l(l+1)}{r^2}+\frac{4m(r)}{r^3} \r]&=&P^t_{\sigma l m} \,,  \nn \\\label{tphicom}
\eea
where we have defined
\be
e^{-u(r)}\equiv 1-2m(r)/r,
\ee
and $m(r)$ is the mass function which denotes the total mass within a sphere of radius $r$. From Eq.~\eqref{bg11}, we get
\be
m(r)=\frac{\kappa}{2}\int_0^r\rho(x)~x^2~dx,
\ee
and the total mass of the star is given by $M \equiv m(r\rightarrow \infty)$. In the equations above, the $P_{\sigma lm}$'s are source terms which depend on the particle's stress-energy tensor, and they are explicitly given, e.g., in Ref.~\cite{Pani:2011xj}. We can also define $h_1(r)$ in terms of the Regge-Wheeler function, 
\be
h_1(r)=-e^{\frac{1}{2}(u-v)}r\Psi_{RW}(r)\,.
\label{h1rw}
\ee
Substituting the relation above into Eq.~\eqref{thetaphicom}, the function $h_0(r)$ can be written in terms of $\Psi_{RW}$ as
\be
h_0(r)=-\frac{i}{\sigma}e^{\frac{1}{2}(v-u)}\frac{d}{dr}\l[r\Psi_{RW}(r)\r]-\frac{i}{\sigma}e^v P_{\sigma l m}(r)\,.
\label{h0rw}
\ee
Equations~\eqref{thetaphicom}-{\eqref{tphicom}} are not all independent, due to the Bianchi identities. Indeed, they are equivalent to a single Regge-Wheeler equation for $\Psi_{RW}$, namely,
\be
\l[\frac{d^2}{dr_*^2} +\sigma^2-V_{RW}(r)\r]\Psi_{RW}(r)=S_{RW}(r)\,,\label{rweq}
\ee
where $r_*$ is the Regge-Wheeler coordinate, defined through $dr_*=e^{(u-v)/2}dr$, $V_{RW}(r)$ is the Regge-Wheeler potential 
\be
V_{RW}(r)=e^v\l[\frac{l(l+1)}{r^2}-\frac{6m(r)}{r^3}-\frac{\kappa}{2}(p_{\text{rad}}-\rho)\r]\,, 
\label{rwpot}
\ee
and $S_{RW}(r)$ is the source term
\be
S_{RW}=\frac{e^{\frac{1}{2}(v-u)}}{r}\l[\frac{2 e^v}{r}\l(1-\frac{rv'}{2}\r)P_{\sigma l m}-e^v P_{\sigma l m}'+P_{\sigma l m}^r\r]\,.\nn
\ee
Note that the homogeneous Regge-Wheeler equation~\eqref{rweq} with the potential~\eqref{rwpot} is equivalent to that of an isotropic, perfect-fluid star with pressure equal to $p_{\rm rad}$~\cite{Kokkotas:1999bd,Thorne1967,Allen:1997xj}.
\subsubsection{Polar sector}\label{polarsec}

The equations for the polar sector are more involved. Following Zerilli~\cite{Zerilli:1971wd}, the linearized Einstein's equations read
\begin{widetext}
\bea
&&K' +\frac{K}{2 r} \left(3-e^{u} \left(1+r^2 \kappa  p_{\text{rad}}\right)\right)+\frac{H_1}{2 r^2} \left(l (l+1)-2 r^2 \kappa (p_{\text{tan}}+\rho)\right)-\frac{H_0}{r}\nn\\
&&+\frac{\kappa}{r^2 \sigma }  \left[r ((\sigma + \omega ) \phi_++(\sigma -\omega) \phi_-) \phi_0'+\omega \phi_0 \left(\phi_+-\phi_--r \phi_+'+r \phi_-'\right)\right]=
\frac{1}{\sigma}A^{(1)}(\sigma,r) - 2 r F(\sigma, r)\,,\label{polareq1}\\
&&H_0'+\frac{K}{2 r} \left(3-e^{u} \left(1+r^2 \kappa  p_{\text{rad}}\right)\right)-\frac{H_0}{r} \left(2-e^{u}(1+ r^2 \kappa  p_{\text{rad}})\right) + \frac{1}{2} H_1 \left(\frac{l(l+1)}{r^2}-2 e^{-v} \sigma ^2-2 \kappa  (p_{\text{tan}}+\rho)\right)\nn\\
&&+\frac{\kappa}{r^2 \sigma }  \left[r ((\omega-\sigma) \phi_+ -(\omega+\sigma) \phi_-) \phi_0'+\omega \phi_0 \left(\phi_+-\phi_--r \phi_+'+r \phi_-'\right)\right] \nn\\
&&=\frac{1}{\sigma}A^{(1)}(\sigma,r)+B(\sigma,r)-r F(\sigma,r)  \left(1-e^{u} \left(1+r^2 \kappa  p_{\text{rad}}\right)\right)\,,\label{polareq2}\\
&& H_1' + (H_0 + K)e^u +\frac{H_1}{r (r-2 m)} \left(2 m-r^3 \kappa  V_0\right)-   \frac{2\kappa }{r \sigma }e^{u}\omega\phi_0 (\phi_+-\phi_-)=
\frac{e^{u}}{ \sigma }B^{(0)}(\sigma,r)+2 r^2 e^u F(\sigma,r)\,,
\label{polareq3}
\eea
\end{widetext}
where the source terms $A^{(1)},~F,~B~$ and $~B^{(0)}~$ read
\bea
A^{(1)}(\sigma,r)&=&\frac{\kappa}{2\sqrt{2}\pi}\int dt~ A^{(1)}_{lm}(r,t) e^{i \sigma t},\\
F(\sigma,r)&=&\frac{\kappa}{2\pi}\sqrt{2\frac{(l-2)!}{(l+2)!}}\int dt~ F_{lm}(r,t) e^{i \sigma t},
\eea
\bea
B(\sigma,r)&=&\frac{\kappa r}{\sqrt{2~ l(l+1)}\pi}\int dt~ B_{lm}(r,t) e^{i \sigma t},\\
B^{(0)}(\sigma,r)&=&\frac{\kappa r}{\sqrt{2~ l(l+1)}\pi} \int dt~ B^{(0)}_{lm}(r,t) e^{i \sigma t}\,,
\eea
and the functions $A^{(1)}_{lm}(r,t)$, $F_{lm}(r,t)$, $B_{lm}(r,t)$ and $B^{(0)}_{lm}(r,t)$ for the Schwarzschild background are explicitly given in Ref.~\cite{Sago:2002fe}. In the background \eqref{metric0}, these functions can be computed in a similar fashion and they reduce to those in Ref.~\cite{Sago:2002fe} in the vacuum case. We have also used that
\be
H_2=H_0-2r^2F(\sigma,r),
\ee
which is obtained from the Einstein equations.  The scalar field perturbations are governed by the following inhomogeneous equations:
\be
\l[\frac{d^2 }{dr_*^2} + (\sigma\pm \omega)^2 -\tilde V\right]\phi_\pm(r)=-\tilde S_\pm\\
\label{eqphiper}
\ee
where
\begin{eqnarray}
 \tilde V&=&e^{v} \left(\frac{l (l+1)}{r^2} + \frac{2m}{r^3} +U_0-\kappa V_0\right)\,,\nn\\
 \tilde S_\pm&=&\frac{e^{-u} \sigma }{2}  \left[2 r (\sigma \pm 2 \omega) \phi_0'\pm \omega \left(4-e^{u} r^2 \kappa  (p_{\text{rad}}+\rho)\right) \phi_0\right]H_1\nn\\
&\pm& r \omega\phi_0 \left[(\sigma\pm 2 \omega ) H_0+ \sigma K + e^{-u} \sigma  H_1'\right]\nn\\
&+&e^{v} r \left[e^{-u} \left(K'-H_0'\right) \phi_0'-(U_0 H_0+\delta {{U}}) \phi_0\right]\nn\\
&+& r^3F \left[\phi_0\left(2 e^{v} U_0-\omega (2 \omega\mp\sigma)\right)+e^{v-u} \phi_0' \left(\frac{2}{r}+\frac{F'}{F}\right)\right]\,.\nn
\end{eqnarray}
Therefore, the polar sector is described by three first-order Einstein equations coupled to two second-order scalar equations. There exists an algebraic relation between $K$, $H_0$ and $H_1$ that can be used to eliminate one of the gravitational perturbations. Finally, the system can be reduced to three coupled second-order differential equations. This is in contrast to the case of perfect-fluid stars, where the polar sector is described by a system of two second-order equations~\cite{Kokkotas:1999bd,PhysRevD.43.1768,Kojima:1992ie}. {Here, rather than working with three second-order equations, we shall use the system of equations given by Eqs. \eqref{polareq1}--\eqref{polareq3} and \eqref{eqphiper}. }
\section{Solving the background equations \label{sec:background}}
In this section we construct spherically symmetric BS models by solving numerically the background equations~\eqref{bg11}--\eqref{bgscalar}. After imposing suitable boundary conditions, the background equations form an eigenvalue problem for the frequency $\omega$, which we solve using a standard shooting method~\cite{Press:1992zz}. 
We integrate Eqs.~\eqref{bg11}--\eqref{bgscalar} from the origin, where we require regularity
\bea
u(r\sim0)&=&0,\\
v(r\sim0)&=&v_c,\\
\phi_0(r\sim0)&=&\phi_c,\\
\phi_0'(r\sim0)&=&0\,.
\eea
The value $v_c$ is arbitrary because it can be adjusted by a time reparametrization in order to impose asymptotic flatness, i.e. $v(r\rightarrow\infty)=0$. In practice, to increase the accuracy of the numerical integration, we have considered a higher order expansion near the origin which, at first order, reduces to the equations above.
At infinity, we impose the metric to be Minkowski and the scalar field to be vanishing:
\be
\phi_0(r\rightarrow \infty)=0\,.
\ee
For each value of $\phi_c$, the boundary condition above is satisfied by a discrete set of eigenfrequencies $\omega$. We focus here on BS background solutions in the ground state, which correspond to the scalar profile having no nodes and to the lowest eigenfrequency $\omega$. The overtones correspond to excited states that would decay to the ground state through emission of scalar and gravitational radiation~\cite{Ferrell:1989kz}.
Note that, depending on the specific BS model, the shooting procedure can be challenging, due to singularities that appear in the integration if the trial frequency $\omega$ is not sufficiently close to the eigenfrequency. In many cases, a precise and tedious fine-tuning is necessary. Furthermore, due to the presence of a mass term in the scalar potential, the scalar field has a Yukawa-like behavior, $(e^{-\sqrt{\mu^2-\omega^2}r_*})/r$ at large distances $r_*\mu\gg 1$~\cite{Schunck:2003kk}. This makes the integration particularly challenging at large distances.

By adopting the procedure above, we can obtain a one-parameter family of solutions, the parameter being the central value of the scalar field $\phi_c$. For each configuration, the total mass of the BS is $M=m(r\to\infty)$. Contrary to the case of perfect-fluid stars, BSs do not possess a well-defined surface as the scalar field spreads all over the radial direction. However, due to the exponential suppression, the configuration is highly localized in a radius $\sim1/\mu$. It is thus useful to define an effective radius for the compact configuration. We shall define the effective radius $R$ such that $m(R)$ corresponds to $99\%$ of the total mass $M$. Other inequivalent definitions have been considered in the literature, see e.g. Ref.~\cite{Schunck:2003kk} for a discussion.

In the following, we describe each of the BS models we have considered, namely, mini BSs, massive BSs and solitonic BSs. 
A summary of the configurations used in this work is presented in Table~\ref{tableconf} (adapted from Ref.~\cite{Macedo:2013qea}). For each BS model, we have selected two stellar configurations. The first configuration corresponds to the maximum total mass of the model, which corresponds to the critical point dividing stable and unstable configurations. The second configuration corresponds to the maximum compactness, defined as $M/R$. Note that the maximum compactness configuration generally occurs for values of $\phi_c$ which are larger than those corresponding to the maximum mass. Therefore, the second configuration is usually in the unstable branch of solutions (cf. e.g. Ref.~\cite{Shapiro:1983du}). 
\begin{table*}
\caption{(Adapted from Ref.~\cite{Macedo:2013qea}) BS models used in this work. For massive BS configurations we use $\tilde\lambda=100$, whereas both solitonic BS models have $\sigma_0=0.05$. The significant digits of $\tilde{\omega}$ do not represent the numerical precision, but they show the fine-tuning needed to achieve the solutions.}
 \begin{tabular}{c | c | c | c | c | c | c }
  \hline\hline
& $\tilde{\phi}_c$ & $\tilde{\omega}$ & $\tilde{M}$ & $\tilde{R}$ & $M\omega $  & $R/M$ \\
 \hline  \hline
mini BS I &0.1916 & 0.853087 & 0.63300 & 7.86149 & 0.54000 &  12.4194 \\
mini BS II & 0.4101 & 0.773453 & 0.53421 & 4.52825 & 0.41319 & 9.03368 \\\hline
massive BS I& 0.094 & 0.82629992558783 &  2.25721 & 15.6565 &    1.86513 &  6.9362 \\
massive BS II &0.155 & 0.79545061700675 &  1.92839 & 11.3739 &    1.53394&   5.8981 \\\hline
  solitonic BS I &1.05 & 0.4868397896964082036868178070 & 1.847287 & 5.72982 &0.89933 & 3.1017\\
solitonic BS II &1.10 & 0.4931624243761699601334882568& 1.698627 & 5.08654 & 0.83770 & 2.9945 \\
\hline\hline
\end{tabular}
\label{tableconf}
\end{table*}
%
\begin{figure*}
\begin{center}
\begin{tabular}{c}
\epsfig{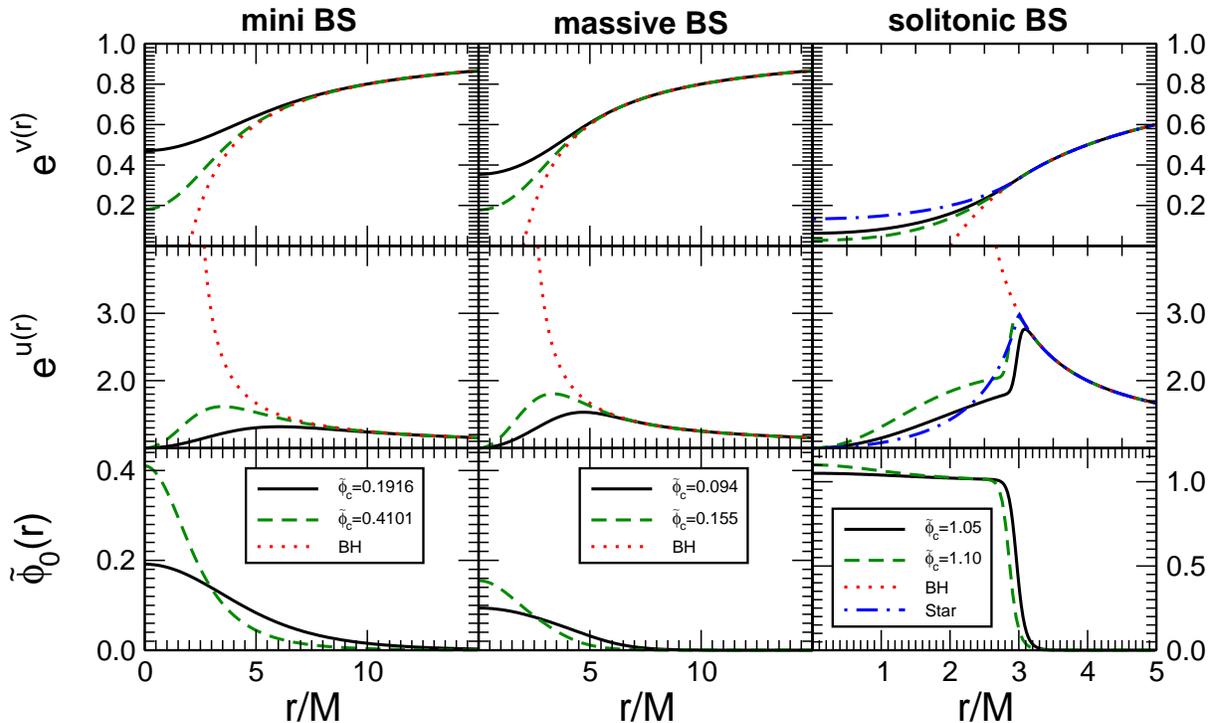}
\end{tabular}
\end{center}
\caption{\label{fig:metric_background}
Rescaled background profiles for different BS models and configurations (cf. Table~\ref{tableconf}). In the top, middle and lower rows we show the metric elements $e^v$, $e^u$ and the scalar profile $\tilde{\phi}_0$, respectively. Each column refers to a different BS model. From left to right: mini BS, massive BS and solitonic BS. For each model, we compare the metric profiles to those of a Schwarzschild BH and for the solitonic BS model we also compare to the metric elements of a uniform density star with $R=3M$.
}
\end{figure*}
%
\subsection{Mini boson stars}
In this model the scalar potential reads
\be
V(|\Phi|^2)=\mu^2|\Phi|^2\,.
\ee
This is one of the simplest potentials that can support self-gravitating configurations. The name comes from the fact that the maximum mass  achieved in this model is smaller than the Chandrasekhar limit for the same particle mass although, for ultralight bosonic fields~\cite{Arvanitaki:2009fg}, it can still reproduce supermassive astrophysical objects. In order to compare with Refs.~\cite{Yoshida:1994xi,Kojima:1991np}, we rescale the equations as
\be
r\to \frac{\tilde{r}}{\mu},\quad m(r)\to \frac{\tilde{m}(\tilde{r})}{\mu},\quad \omega \to \tilde{\omega}\mu,\quad \phi_0 (r) \to \frac{\tilde{\phi}_0(\tilde{r})}{\sqrt{4\pi}}. \nn
\ee
The rescaled background profiles (metric functions and the scalar field) for the two configurations listed in Table~\ref{tableconf} are shown in the left panels of Fig.~\ref{fig:metric_background}. The metric functions for these configurations are also compared with the Schwarzschild black hole ones.

\subsection{Massive boson stars}
For this model the potential has a quartic interaction:
\be
V(|\Phi|^2)=\mu^2|\Phi|^2+\frac{\lambda}{2}|\Phi|^4\,,
\ee
where $\lambda$ is a constant. This potential was studied in Ref.~\cite{Colpi:1986ye}, where it was shown that the model may differ considerably from the mini BS case, even when $\lambda \ll 1$. Also, the maximum mass increases with $\lambda$, being comparable with the Chandrasekhar limit. For the case of massive BSs, in order to facilitate the comparison with the results in Ref.~\cite{Colpi:1986ye}, we have performed the following rescaling:
\bea
&&r\to \frac{\tilde{r}}{\mu},\quad
m(r)\to \frac{\tilde{m}(\tilde{r})}{\mu},\quad
\omega \to \tilde{\omega}\mu,\quad\nn\\
&&\lambda \to 8\pi \mu^2 \tilde\lambda,\quad
\phi_0 (r) \to \frac{1}{2\sqrt{2\pi}} \tilde{\phi}_0(\tilde{r})\,.
\eea
The maximum compactness for solutions of this model increases with $\lambda$, and we found results in agreement with previous calculations~\cite{Guzman:2005bs,AmaroSeoane:2010qx}. Here, we fixed $\tilde\lambda=100$ and considered two configurations as summarized in Table~\ref{tableconf}. The metric and scalar field profiles for this model are shown in the middle panels of Fig.~\ref{fig:metric_background}.

\subsection{Solitonic boson stars}
The scalar potential for this configuration is given by
\be
V(|\Phi|^2)=\mu^2|\Phi|^2 (1-2|\Phi|^2/\sigma_0^2)^2\,, \label{potsolitonic BSs}
\ee
where $\sigma_0$ is a constant, generically taken to be of the same order as $\mu$~\cite{Friedberg:1986tq,Lee:1991ax}. This is the simplest potential that can generate, in the absence of gravity, nontopological solitonic solutions, i.e., nondispersive scalar field solutions.
In this case, it is convenient to rescale the equations in units of $\Lambda \mu$, with $\Lambda = \kappa^{1/2}\sigma_0$. We use~\cite{Friedberg:1986tq,Kesden:2004qx}
\bea
&&r\to \frac{\tilde{r}}{\Lambda \mu},\quad
m(r)\to \frac{\tilde{m}(\tilde{r})}{\Lambda \mu},\nn\\
&&\omega \to \tilde{\omega} \Lambda \mu,\quad
\phi_0 (r) \to \frac{\sigma_0 \tilde{\phi}_0(\tilde{r})}{\sqrt{2}}\,.\nn
\eea
The field equations for the solitonic potential are stiff, and the scalar field has a very steep profile across a surface layer of thickness $\sim\mu^{-1}$. This stiffness makes the numerical integration particularly challenging and, in Refs.~\cite{Friedberg:1986tq,Kesden:2004qx}, spherically symmetric solutions to this model were constructed only perturbatively, in the limit $\sigma_0\ll m_{\rm P}$ and considering a step-function profile for the scalar field. One advantage of that approach is that the approximate solution has a well-defined radius and that, because the scalar profile is given, only the metric equations have to be solved numerically in the interior of the star. The solution is then matched with a Schwarzschild exterior.

However, besides the challenging technicalities in the integration, there is no real need to obtain approximate solutions, which neglect the backreaction between metric functions and the scalar field. Here, we have constructed solitonic BS solutions to the \emph{full} nonlinear system~\eqref{bg11}--\eqref{bgscalar}, i.e. without any approximation (cf. also Refs.~\cite{Kleihaus:2005me} where similar solutions were constructed using relaxation methods). This requires high-precision numerical schemes and an extremely fine-tuned shooting method, as shown by the fine-tuning needed to find a solution (cf. Table~\ref{tableconf}). In the small $\sigma_0$ limit, our results agree remarkably well with the approximate solutions presented in Refs.~\cite{Friedberg:1986tq,Kesden:2004qx} and they extend those results to generic values of the parameters in the scalar potential~\eqref{potsolitonic BSs}.

Unlike the other cases explored in this paper, solitonic BSs can be very compact, with the radius of the star comparable to or smaller than the Schwarzschild light ring~\cite{Friedberg:1986tq,Kesden:2004qx}. In the right panels of Fig.~\ref{fig:metric_background} we compare the metric components to those of a Schwarzschild spacetime and of the uniform density stars with $R=3M$, and we show the steep profile of the scalar field. The scalar field approximates a step function, in agreement with the approximate solution of Refs.~\cite{Friedberg:1986tq,Kesden:2004qx}. In that case $e^{u(r)}$ is discontinuous at the star surface. In our case there is no actual radius, and $e^{u(r)}$ is continuous, although it has a sharp peak close to the effective radius of the star.

\section{Geodesics around boson stars \label{sec:geodesics}}
%
\begin{figure*}
\begin{center}
\begin{tabular}{c}
\epsfig{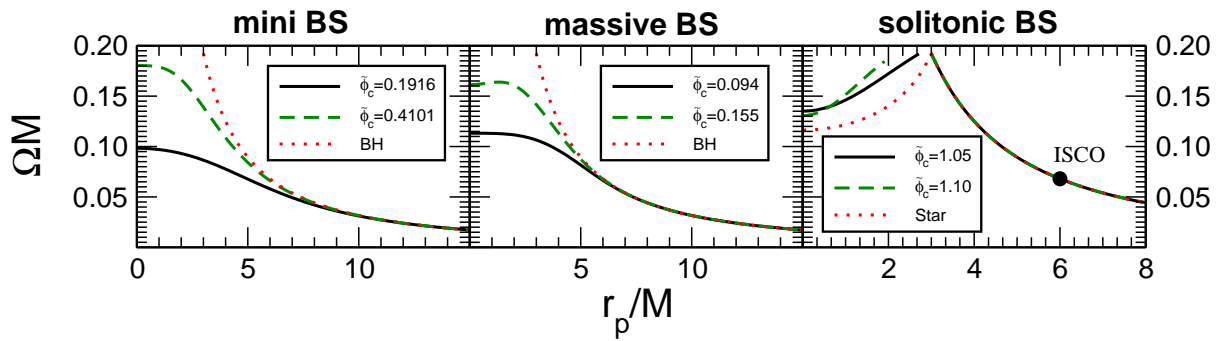}
\end{tabular}
\end{center}
\caption{\label{fig:circular_geodesics} (Adapted from Ref.~\cite{Macedo:2013qea})
Angular velocity for timelike circular geodesic motion for the different BS models and configurations specified in Table~\ref{tableconf}. Each plot refers to a different BS model. From left to right: mini BS, massive BS and solitonic BS. For mini and massive BSs we compare the angular velocity to those of a Schwarzschild BH, and for the solitonic BS model we compare to the case of a uniform density star with $R=3M$. In the solitonic case, the marker indicates the innermost stable circular orbit for the Schwarzschild BH, which is given by $r=6M$ and $M\Omega_{isco}\approx 0.068$.
}
\end{figure*}

Stellar-size objects gravitating around supermassive BSs have a small backreaction on the geometry
and, to first order in the object's mass, move along geodesics of the BS background.
Accordingly, gravitational-wave emission by such binaries requires a knowledge of the geodesic motion, on which we now focus.
We will also concentrate exclusively on circular, geodesic motion. The reasoning behind this is that it makes the calculations much simpler,
while retaining the main features of the physics. Furthermore, it can be shown that generic eccentric orbits get circularized by gravitational-wave emission in vacuum \cite{Peters:1964zz} and in the presence of accretion and gravitational drag \cite{Macedo:2013qea}, on a time scale that depends on the mass ratio. 

We follow the analysis by Chandrasekhar~\cite{Chandrasekhar:1985kt} (see also Ref.~\cite{Cardoso:2008bp}, where the formalism for a generic background is presented, and Ref.~\cite{Eilers:2013lla} for a recent work on geodesics in BS spacetimes). Following previous studies~\cite{Kesden:2004qx,Guzman:2005bs,Torres:2002td}, we assume that the point particle is not directly coupled to the background scalar field.
We start by defining the Lagrangian of the particle motion on the $\theta=\pi/2$ plane:
\be
2 \mathcal{L}_p = \dot{s}^2=-e^v \dot{t}^2+e^u \dot{r}^2+r^2 \dot{\varphi}^2\,.
\ee
The conserved energy $E$ and angular-momentum parameter per unit rest mass $L$ and can be obtained via
\be
E=-\frac{\pa \mathcal{L}_p }{\pa \dot{t}}=e^v \dot{t}\,,\qquad
L=\frac{\pa \mathcal{L}_p }{\pa \dot{\varphi}}=r^2 \dot{\varphi}\,.
\ee
From these equations, we get the following equation of motion:
\be
e^{u+v}\dot{r}^2=E^2-V_{eff}(r)=E^2-e^v\l(1+\frac{L^2}{r^2}\r)\,.
\label{balance}
\ee
The energy and angular-momentum of the particle in circular orbits follow from Eq.~\eqref{balance} by imposing $\l.\dot{r}\r|_{r=r_p}=0$ and $\l.\ddot{r}\r|_{r=r_p}=0$, resulting in
\bea
E_c=\l[ e^{v}\frac{2 (r-2 m)}{2 r-\kappa\, r^3\, p_{\text{rad}}-6 m}\r]_{r=r_p}^{1/2}\\
L_c=\l[r^2 \frac{\left(\kappa \,r^3\, p_{\text{rad}}+2 m\right)}{2 r-\kappa\, r^3 \,p_{\text{rad}}-6 m}\r]_{r=r_p}^{1/2},
\eea
where the background Einstein equations were used to eliminate metric derivatives. Circular null geodesics correspond to $2 r-\kappa r^3 p_{\text{rad}}-6 m=0$. Finally, the orbital frequency of circular geodesics reads
\be
\Omega=\l[\frac{e^{v} \left(\kappa \,r^3\, p_\text{rad}+2 m\right)}{2 r^2 (r-2 m)}\r]_{r=r_p}^{1/2}.
\ee

The angular velocities of circular geodesics in BS spacetimes are shown in Fig.~\ref{fig:circular_geodesics}.
Up to the innermost stable circular orbit of a Schwarzschild spacetime, $r=6M$, the angular velocities are very close to their Schwarzschild counterpart with the same total mass, as might be expected since these are very compact configurations. For geodesics at $r<6M$ the structure can be very different. A striking difference is that stable circular timelike geodesics exist for BSs even very deep into the star~\cite{Torres:2002td,Guzman:2005bs,Macedo:2013qea}.

Solitonic BSs can become truly relativistic gravitating objects. For these objects, an outer last stable circular orbit exists at $r\approx 6M$ and $M\Omega_{isco}\approx 0.068$. This is expected, as the spacetime is very close to Schwarzschild spacetime outside the solitonic BS effective radius. We also find a first (unstable) light ring at roughly $r_{l_+}\approx3M$.
The unexpected feature is the presence of a second {\it stable} light ring at $r_{l-}<r_{l_+}$, together with a family of stable timelike circular geodesics all the way to the center of the star. These light rings are genuine relativistic features, which was not reported in previous studies, as far as we are aware.
Uniform density stars, depending on their compactness, also present two light ring and stable circular timelike orbits in their interior.
In the right panel of Fig.~\ref{fig:circular_geodesics}  the case of a uniform density star with radius $R=3M$ is also shown. In that case, the two light rings degenerate in the star surface. What makes solitonic BSs stand out is the possibility that inspiraling matter couples weakly to the solitonic BS scalar field and therefore has access to these geodesics, although as we showed in Ref.~\cite{Macedo:2013qea}, inspiraling BHs in principle do {\it not} follow these geodesics.
Furthermore, we found no circular orbits between the outer and the inner light ring, whereas all circular orbits are stable inside the inner light ring.

Finally, deep inside the BSs, the circular geodesics are nonrelativistic. In fact, the velocity as measured by static observers at infinity and by static observers at fixed $r$, decreases to zero as the radius approaches zero. In this regime, other dissipative effects such as gravitational drag and accretion onto the small compact object have to be considered~\cite{Macedo:2013qea}.

\section{Quasinormal modes of boson stars \label{sec:qnms}}
In this section we discuss the quasinormal modes (QNMs) of the BS models presented in the previous sections.
QNMs are complex eigenfrequencies $\sigma=\sigma_R+i\sigma_I$ of the linearized homogeneous perturbation equations supplied with physically motivated boundary conditions (see e.g.~\cite{Kokkotas:1999bd,Berti:2009kk}). Since the perturbations of a spherically symmetric spacetime naturally divide into an axial and a polar sector, there exist two different classes of modes, which we shall refer to as axial and polar modes, respectively.

Unlike the case of a Schwarzschild BH~\cite{Chandrasekhar:1985kt}, the axial and the polar BS modes are not isospectral. As we shall discuss, the BS QNMs can be understood in analogy to the modes of ordinary stars, with the background scalar field playing the role of an anisotropic fluid. The main difference with the case of ordinary stars is that a BS does not have a proper surface and that scalar perturbations, unlike their fluid counterpart, can propagate to infinity. 
In the following, we shall treat axial and polar modes separately.
\subsection{Axial QNMs}
\begin{figure*}
\begin{center}
\begin{tabular}{c}
\epsfig{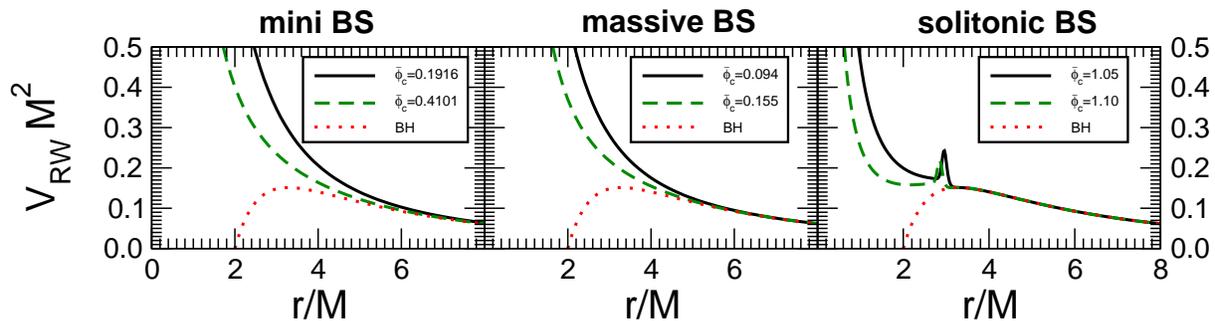}
\end{tabular}
\end{center}
\caption{\label{fig:rwpotential}Regge-Wheeler potential for the mini BS, massive BS and solitonic BS models compared to that of the Schwarzschild BH for $l=2$.
}
\end{figure*}

As discussed in Sec.~\ref{axials}, the source-free ($S_{RW}=0$) axial perturbations can be reduced to the homogeneous Regge-Wheeler equation
\be
\l[\frac{d^2}{dr_*^2} +\sigma^2-V_{RW}(r)\r]\Psi_{RW}(r)=0\,,
\label{rweqh}
\ee
where $V_{RW}$ is defined in Eq.~\eqref{rwpot} and it is shown in Fig.~\ref{fig:rwpotential} for some BS model and for the case of a Schwarzschild BH.
Note that Eq. \eqref{rweqh} does not involve scalar field perturbations, in analogy to the fluid perturbations of an ordinary star, which are only coupled to the polar sector. This decoupling led Yoshida\textit{ et al.}~\cite{Yoshida:1994xi} to assume that the axial sector of BSs is ``not coupled to gravitational waves'' and therefore not interesting. However, we show here that BS models generically admit axial QNMs, in analogy to the $w$ modes of ordinary stars which are in fact curvature modes similar to those of a BH (see Ref.~\cite{Kokkotas:1999bd} for a review). Moreover, for ultracompact stars ($R<3M$), a potential well appears in the Regge-Wheeler potential, generating the possibility of having trapped QNMs, which are long-living modes~\cite{chandrasekharferrariIII,Kokkotas:1994an}. In Sec.~\ref{sec:PP} we shall also show that axial perturbations with odd values of $l+m$ are sourced by point particles orbiting the BS, and therefore they contribute to the gravitational-wave signal emitted during the inspiral.

At the center of the star, we require regularity of the Regge-Wheeler function, 
\be
\Psi_{RW}(r\approx 0) \sim r^{l+1}\sum_{i=0}^Na^{(i)}_0 r^i\,,\label{BCaxial0}
\ee
where the coefficients $a^{(i)}_0$ can be obtained by solving the Regge-Wheeler equation order by order near the origin. At infinity, the solution of Eq.~\eqref{rweqh} is a superposition of ingoing and outgoing waves. The QNMs are defined by requiring purely outgoing waves at infinity, i.e.
\be
\Psi_{RW}(r\rightarrow \infty) \sim e^{i \sigma r_*}\sum_{i=0}^N\frac{a^{(i)}_\infty}{r^i}\, ,
\label{BCaxialINF}
\ee
where again the coefficients $a^{(i)}_\infty$ can be obtained perturbatively. In the following we discuss two different methods to compute BS axial modes.
\subsubsection{Axial QNMs via continued fractions}
In Ref.~\cite{Yoshida:1994xi}, the polar modes of some mini BS configurations were computed using a WKB approximation. Here, we resort to a continued fraction method~\cite{Leaver:1985ax} adapted from the studies of ordinary stars as shown in Refs.~\cite{Leins:1993zz,Benhar:1998au} (see also Ref.~\cite{Pani:2009ss} in which the same method was applied to gravastars).

First, we write the solution of the homogeneous Regge-Wheeler equation in a power-series expansion of the form 
\be
\Psi_{RW}(r)=(r-2M)^{2iM\sigma}e^{i\sigma r}\sum_{n=0}^\infty a_n z^n\,,\label{subst}
\ee
where $z\equiv 1-R_2/r$, and $r=R_2$ is some point outside the stellar object (in our case will be outside the effective radius). The expansion coefficients $ a_n$ are found to satisfy a four-term recurrence relation of
the form:
\beq
\label{4trrgiusta}
&&\alpha_1 a_{2}+\beta_1 a_1+\gamma_1 a_{0}=0\,, \quad n=1\,,\\
&&\alpha_n a_{n+1}+\beta_n a_n+\gamma_n a_{n-1}+\delta_n a_{n-2}=0\,,
\quad n\geq 2\,,
\nonumber
\eeq
where:
\beq\label{primostep1}
&&\alpha_n=n(n+1)(R_2-2M)\,,\quad n\geq 1\,,\\
\nn
&&\beta_n=2 n (-3 M n+R_2 (n-iR_2 \sigma ))\,,\quad n\geq 1\,,\\
\nn
&&\gamma_n=6 M ((n-1) n-1+)+(1+l-n) (l+n)R_2\,,\\
\nn
&&\delta_n=2 M (3-n) (1+n)\,,\quad n\geq 2\,.
\eeq
Since the Regge-Wheeler equation is homogeneous, the coefficient $a_0$ is an arbitrary normalization constant. The ratio
$a_1/a_0$ can be determined by imposing the continuity of $ \Psi_{RW}$ and
$\Psi_{RW}'$ at $r=R_2$. From Eq.~\eqref{subst} it follows that:
\be
\frac{a_1}{a_0}=\frac{R_2}{\Psi_{RW}(R_2)}
\left[\Psi_{RW}'(R_2)-\frac{i\sigma R_2}{R_2-2M}\Psi_{RW}(R_2)\right]\,.
\label{a1a0}
\ee
As in the case of ordinary stars, the values of $\Psi_{RW}(R_2)$ and $\Psi_{RW}'(R_2)$ are obtained by integrating numerically
the Regge-Wheeler equation in the interior. Leaver~\cite{Leaver:1990zz} has shown that the four-term
recurrence relation~\eqref{4trrgiusta} can be reduced to a three-term
recurrence relation by a Gaussian elimination step and solved by standard methods~\cite{Berti:2009kk} (see also Ref.~\cite{Pani:2009ss} for a more detailed discussion). The complex roots of the continued fraction relation are the QNMs of the BS.  
\subsubsection{Axial QNMs via direct integration}
In some cases, QNMs can be computed via direct integration~\cite{Chandrasekhar:1975zza,Pani:2013pma}. This method is not particularly well suited because radial QNM functions grow exponentially as $r\to\infty$ and become very sensitive to numerical errors~\cite{Berti:2009kk}.
However, it is possible to integrate Eq.~\eqref{rweqh} up to moderately large values of $r$ and to minimize the truncation errors by considering a large number of terms in the series expansion~\eqref{BCaxialINF}. In our code, we typically consider $N=15$ in Eq.~\eqref{BCaxialINF} and integrate up to $r\sim 30 M$. This would suppress truncation errors at the level of $30^{-15}\sim 10^{-22}$.

The method is a simple extension of the case of uniform density stars~\cite{Chandrasekhar:1975zza,Kokkotas:1994an,chandrasekharferrariIII}. We perform two integrations of Eq.~\eqref{rweqh}: one from the center of the star  with the boundary condition~\eqref{BCaxial0} up to $r_{m}$, and another from $r_{\infty}$ with the boundary condition~\eqref{BCaxialINF} until $r_m$. The wave functions constructed this way have the correct boundary conditions both at the origin and at infinity. However, for generic values of the frequency $\sigma$ the Regge-Wheeler function is not continuous at the matching point $r=r_m$. We define the jump at $r_m$ as~\cite{Chandrasekhar:1975zza}
\be
\Delta_m(\sigma)\equiv \l[\frac{d\Psi_{RW}/{dr_*}}{\Psi_{RW}}\r]_{-} -\l[\frac{d\Psi_{RW}/{dr_*}}{\Psi_{RW}}\r]_{+}\,,
\ee
where the ``minus'' and ``plus'' subscripts denote evaluation at $r=r_m$ from the left and from the right, respectively.
The axial QNMs are obtained as the roots of $\Delta_m(\sigma)$. Due to the numerical inaccuracies discussed above, this procedure becomes less accurate for modes with large imaginary part. For example it can be used to obtain only the first few tones of a Schwarzschild BH~\cite{Chandrasekhar:1975zza}.

A similar procedure can be adopted in the case of ordinary stars, this time by requiring that the Wronskian of the two solutions (those constructed by integrating from the center and from infinity) is vanishing at the star surface. This is equivalent to requiring continuity of the wave function and of its first derivative. 
To test our code, we successfully found some of the modes presented in Refs.~\cite{Kokkotas:1994an,chandrasekharferrariIII} for constant density stars, whose background metric coefficients can be determined analytically (see Appendix~\ref{app:stars}). 
For all modes computed by direct integration, we have checked the stability of the results under variation of the parameters $r_m$ and $r_\infty$.
We stress that, at variance with continued fraction techniques, the direct integration is only accurate when $\sigma_I\ll\sigma_R$.

\subsection{Polar QNMs}
As discussed in Sec.~\ref{polarsec}, the polar sector can be reduced to a system of three coupled second-order differential equations: 
two for the scalar field perturbations $\phi_\pm$ and one for gravitational perturbations described by a modified Zerilli equation. 
In practice, in the interior of the object it is more convenient to solve directly for the polar perturbation functions, $K$, $H_0$ and $H_1$, which are described by three first-order differential equations and by an algebraic relation.

As in the axial case, at the origin we require regularity of the perturbations and we can expand them in powers of $r$ as
\begin{equation}
 X(r\approx 0)\sim r^l \sum_{i=0}^N x_0^i~ r^{i}\,,
\end{equation}
where $X$ collectively denotes $H_2=H_0$, $K$, $H_1$ and $\phi_\pm$.
It is straightforward to show that this expansion near the center only depends on three free parameters. 

At infinity, the background scalar field vanishes and gravitational and scalar perturbations decouple~\cite{Yoshida:1994xi}. 

Let us now discuss the asymptotic behavior of the gravitational field. In vacuum, all polar metric perturbations can be written in terms of one single function which obeys the Zerilli equation,
\be
\l[\frac{d^2}{dr_*^2}+\sigma^2-V_Z(r)\r]\Psi_{Z}=0\,,
\ee
where $dr/dr_*=1-2M/r$,
\be
V_Z(r)=\frac{dr}{dr_*}\frac{2 \tilde\Lambda ^2r^2(3M +(\tilde\Lambda +1)r)+18M^2(\tilde\Lambda r+M)}{r^3(\tilde\Lambda  r+3M)^2}\,,
\ee
and $\tilde\Lambda =(l-1)(l+2)/2$. The generic solution at infinity is a superposition of outgoing and incoming waves:
\be
\Psi_{Z}(r\rightarrow \infty)\sim A_{\rm out} e^{i \sigma r_*}+A_{\rm in} e^{-i \sigma r_*} 
\,,
\label{bounzer}
\ee
and again the standard QNM condition requires $A_{\rm in}=0$~\cite{Berti:2009kk}.
The metric perturbations can be written in terms of the Zerilli function through the following equations:
\bea
H_1&=&-\frac{\tilde\Lambda  r^2-3\tilde\Lambda  M r -3 M^2}{(r-2M)(\tilde\Lambda  r+3M)}\Psi_Z-r^2\frac{d\Psi_Z/dr_*}{r-2M}\,,\nn\\
K&=&\frac{\tilde\Lambda (\tilde\Lambda +1)r^2+3M \tilde\Lambda  r+6M^2}{r^2(\tilde\Lambda  r+3M)}\Psi_Z+\frac{d\Psi_Z}{dr_*}\,,\nn\\
H_0&=&H_2=\frac{\tilde\Lambda  r (r-2 M)-\sigma ^2 r^4+M(r-3M) }{(r-2M)(\tilde\Lambda  r+3 M)}K\nn\\
&&+\frac{M(\tilde\Lambda +1)-\sigma ^2r^3}{r(\tilde\Lambda  r+3M)}H_1\,.\nn
\eea

The asymptotic behavior of the scalar field perturbations is more involved. In vacuum, the equations for the scalar perturbations~\eqref{eqphiper} reduce to 
\be
\l[\frac{d^2}{dr_*^2}+(\sigma \pm \omega)^2-V_\phi(r)\r]\phi_{\pm}=0\,,
\ee
where
\be
V_\phi(r)=\l(1-\frac{2M}{r}\r)\l(\mu^2+\frac{l(l+1)}{r^2}+\frac{2M}{r^3}\r)\,.
\ee
The asymptotic solution for the scalar perturbations reads
\be
\phi_\pm(r\rightarrow \infty) \sim B_\pm e^{-k_\pm r_*}r^{\nu_\pm}+C_\pm e^{k_\pm r_*}r^{-\nu_\pm} 
\,,
\ee
where we have defined $\nu_\pm={M\mu^2}/{k_\pm}$ and
\be
k_{\pm}=\sqrt{\mu^2-(\sigma \pm \omega)^2}\,. \label{kpm}
\ee
Without loss of generality, we choose the root such that ${\rm Re}[k_\pm]>0$.
Different physically motivated boundary conditions are possible for the scalar field, depending on the sign of the imaginary part of $k_\pm$, ${\rm Im}[k_\pm]\sim-\left(\sigma_R\pm\omega\right)\sigma_I$. As usual, a purely outgoing-wave boundary condition at infinity, i.e. $\phi_\pm\sim e^{i|{\rm Im}[k_\pm]|r_*}$, defines the QNMs. On the other hand, due to the presence of the mass term it is possible to have quasibound-state modes, i.e. states that are spatially localized within the
vicinity of the compact object and decay exponentially at spatial infinity~\cite{Dolan:2007mj,Rosa:2011my,Pani:2012bp}. 
Therefore, quasibound states are simply defined by $C_\pm=0$. 
In the case at hand, the QNM conditions depend on $\sigma_R$ and on $\sigma_I$, as shown in Table~\ref{table:BCs} where all cases are listed. In the following, we detail the QNM condition for stable and unstable modes.

Let us start discussing the boundary conditions for stable modes ($\sigma_I<0$). When $\sigma_R>\omega$ the QNM condition is the same for both scalar perturbations, $B_\pm=0$.
However, if $\sigma_R<\omega$, the QNM condition for the scalar field perturbations is different, being $B_+=0$ and $C_-=0$. Note that in this case the stable QNMs of $\phi_-$ decay exponentially and is degenerate with the bound-state modes.

For unstable modes ($\sigma_I>0$) the situation is different. In this case when $\sigma_R>\omega$, the QNM condition is the same for both scalar perturbations, $C_\pm=0$, and coincide with the bound-state conditions. However, when $\sigma_R<\omega$ the QNM conditions read $C_+=0$ and $B_-=0$, so that only the unstable QNM condition of $\phi_+$ coincides with the bound-state condition.

This peculiar behavior is due to the presence of a mass term (which allows for bound states) \emph{and} of a complex background scalar field, $\omega\neq0$, which essentially shifts the real part of the frequency of the scalar perturbations. Note that in the case of probe complex scalars around a Schwarzschild BH, the terms introduced by $\omega$ can be eliminated by a simple shift of the wave frequency, but in the case at hand, this term is physical because of the coupling to the gravitational perturbations.
\begin{table*}
\caption{Possible boundary conditions at infinity for the scalar field perturbations $\phi_\pm$ with eigenfrequency $\sigma=\sigma_R+i\sigma_I$.}
 \begin{tabular}{c | c | c | c | c | c  }
  \hline\hline
  &$\sigma_I$		& $\sigma_R$		& ${\rm Im}[k_\pm]$			& QNM condition		& Bound-state condition\\
 \hline  \hline
I &Stable, $\sigma_I<0$	&$\sigma_R>\omega$	&${\rm Im}[k_+]>0$, ${\rm Im}[k_-]>0$	&$B_+=0$, $B_-=0$	&$C_+=0$, $C_-=0$ \\	
II &Stable, $\sigma_I<0$	&$\sigma_R<\omega$	&${\rm Im}[k_+]>0$, ${\rm Im}[k_-]<0$	&$B_+=0$, $C_-=0$	&$C_+=0$, $C_-=0$ \\	
\hline
III &Unstable, $\sigma_I>0$	&$\sigma_R>\omega$	&${\rm Im}[k_+]<0$, ${\rm Im}[k_-]<0$	&$C_+=0$, $C_-=0$	&$C_+=0$, $C_-=0$ \\	
IV &Unstable, $\sigma_I>0$	&$\sigma_R<\omega$	&${\rm Im}[k_+]<0$, ${\rm Im}[k_-]>0$	&$C_+=0$, $B_-=0$	&$C_+=0$, $C_-=0$ \\	
\hline\hline
\end{tabular}
\label{table:BCs}
\end{table*}
%

\subsubsection{Polar QNMs via direct integration}
Computing the polar modes of a BS is particularly challenging. To compute the polar QNMs of perfect fluid stars the usual continued fraction method proves to be very robust. However, unlike the case of ordinary stars, BSs do not possess a surface where fluid perturbations vanish.
In order to understand this issue, let us briefly review the case of ordinary stars~\cite{Leins:1993zz,Benhar:1998au}. In that case polar QNMs are found by first solving a boundary problem in the interior of the star, requiring the perturbations to be regular at the center and the pressure perturbations to be vanishing at the surface of the star. For any given frequency, this procedure singles out one solution that satisfies the correct boundary condition in the interior, and it allows one to construct the Zerilli function $\Psi_Z$ at the radius of the star. Then, Chandrasekhar transformations~\cite{Chandrasekhar:1985kt} are used to transform the Zerilli function into the Regge-Wheeler function $\Psi_{RW}$ and, finally, the continued fraction method can be implemented as explained above for the axial case.

Contrarily to the case of fluid perturbations in ordinary stars, in the BS cases the matter perturbations (scalar field perturbations) propagate in vacuum and, strictly speaking, there is no exterior Schwarzschild solution in which the linear dynamics is simply governed by a single Regge-Wheeler equation. This prevents a direct extension of this method.

To circumvent this problem, we opt for direct integration techniques, which we now describe. The system of linearized perturbation equations can be written as a first-order system for the six-dimensional vector $\mathbf{\Psi}=\l(H_1,K,\phi_+,\phi_-,\phi_+',\phi_-'\r)$. We perform two integrations: one from the origin and one from infinity, in both cases imposing suitable boundary conditions as discussed above. It is easy to show that, for each integration, there exists a three-parameter family of solutions, corresponding to three independent parameters of the near-origin and near-infinity expansions. Then, we construct the linear combinations
\begin{eqnarray}
 \mathbf{\Psi}_-&=&\alpha^{(-)}_1 \mathbf{\Psi}_1^{(-)}+ \alpha^{(-)}_2 \mathbf{\Psi}_2^{(-)} + \alpha^{(-)}_3 \mathbf{\Psi}_3^{(-)}\,,\\
 \mathbf{\Psi}_+&=&\alpha^{(+)}_1 \mathbf{\Psi}_1^{(+)}+ \alpha^{(+)}_2 \mathbf{\Psi}_2^{(+)} + \alpha^{(+)}_3 \mathbf{\Psi}_3^{(+)}\,,
\end{eqnarray}
where $\alpha_i^{(\pm)}$ are constants and $\mathbf{\Psi}_-$ and $\mathbf{\Psi}_+$ refer to the integration from the origin and from infinity, respectively. The subscripts $1$, $2$ and $3$ refer to three linear independent solutions of the homogeneous system. Since the system of equations is linear, we have the freedom to set one of the coefficients $\alpha_i^{(\pm)}$ of the linear combination to unity. The other five coefficients can be obtained by requiring $\mathbf{\Psi}_-=\mathbf{\Psi}_+$ at some arbitrary matching point. For a generic frequency, only five out of the six components of $\mathbf{\Psi}$ can be matched smoothly. Finally, the eigenfrequency of the problem is obtained by requiring that the remaining component is also	 continuous. In practice, for each frequency $\sigma$ we can perform six numerical integrations of the linear system, construct the linear combinations above, obtain the coefficients $\alpha_i^{(\pm)}$ and compute the jump of the only discontinuous component of $\mathbf{\Psi}$ at the matching point. Then, a standard shooting method can be implemented to obtain the complex eigenfrequency. Similarly to the direct integration discussed in the axial case, this method provides accurate results only when $\sigma_I\ll\sigma_R$.

\subsection{Results for BS QNMs}
Using the methods described above, we have computed axial and polar modes of several BS models in a fully relativistic setting, i.e. without using any approximation method. As shown in Table~\ref{table:BCs}, the spectrum of BS polar modes is fairly rich. 
Here, we focus on the least damped modes, i.e. those with the smallest imaginary part, which are expected to dominate the ringdown waveform at late times~\cite{Berti:2009kk}. Note that, for all BS models we have investigated, there exists a class of much longer lived modes than that considered in Ref.~\cite{Yoshida:1994xi}. We have also shown the modes in units of $M$, for future comparisons. In the tables, $N\geq1$ is the overtone number.

For the axial modes, we have used the continued fraction method and, for the modes with $\sigma_I\ll \sigma_R$, we independently confirmed the results by using a direct integration method. The direct integration works better for compact configurations like the solitonic BSs, which share many similarities with compact uniform density stars. The least damped axial QNMs of solitonic BSs are presented in Table~\ref{tableqnsaxialbo}, comparing the results of the two different methods. 
\begin{table*}
\caption{Axial QNMs for solitonic BS configurations~I and II for $l=1$ and $l=2$. Here we compare the results obtained through the continued fraction and the direct integration methods.}\label{tableqnsaxialbo}
$l=1$\\
Continued fraction\hspace{6cm}Direct integration\\
 \begin{tabular}{| c | c | c | c | c | c |}
\hline
 Model & N & Re$(\sigma)~ [\Lambda^2 \mu]$ & -Im$(\sigma) ~[\Lambda^2 \mu]$ & Re$(M\sigma)$ & -Im$(M\sigma)$ \\
\hline
Solitonic BS I & 1 & 0.22328867 & 0.08370555 & 0.412478 & 0.154628 \\
Solitonic BS I & 2 & 0.38509593 & 0.10287792 & 0.711383 & 0.190045\\
Solitonic BS I & 3 & 0.55353269 & 0.11432831 & 1.022530  & 0.211197\\
\hline
Solitonic BS II & 1 & 0.20007784 & 0.06608236 &  0.339858   & 0.112249\\
Solitonic BS II & 2 & 0.32840222 & 0.08229700 &  0.557833   & 0.139792\\
Solitonic BS II & 3 & 0.46744415 & 0.09216367 &  0.794014   & 0.156552\\
\hline 
 \end{tabular}
 \begin{tabular}{| c | c |}
\hline
 Re$(\sigma)~ [\Lambda^2 \mu]$ & -Im$(\sigma) ~[\Lambda^2 \mu]$  \\
\hline
0.22329050 &  0.08370789  \\
0.38509593 &  0.10287792 \\
0.55353269 &  0.11432831\\
\hline
0.20007454 &  0.06608373 \\
0.32840223 &  0.08229700\\
0.46744415 & 	0.09216367 \\
\hline
 \end{tabular}
\\
\vspace{.2cm}
$l=2$\\
\vspace{.1cm}
 \begin{tabular}{| c | c | c | c | c | c |}
\hline
 Model & N & Re$(\sigma)~ [\Lambda^2 \mu]$ & -Im$(\sigma) ~[\Lambda^2 \mu]$ & Re$(M\sigma)$ & -Im$(M\sigma)$ \\
\hline
Solitonic BS I & 1 & 0.25636868 & 0.05347247 & 0.47358 & 0.098779 \\
Solitonic BS I & 2 & 0.32633835  &0.10252772  & 0.60284 & 0.189398\\
Solitonic BS I & 3 &0.47822011  & 0.10629265 & 0.88341 & 0.196353\\
\hline
Solitonic BS II & 1 &  0.26620716 & 0.02511717 &  0.452187 & 0.0426647\\
Solitonic BS II & 2 &  0.32967926 & 0.08729943& 0.560002 & 0.148289\\
Solitonic BS II & 3 & 0.41859619 &  0.08681748 &  0.711039 & 0.147471\\
\hline 
 \end{tabular}
 \begin{tabular}{| c | c |}
\hline
 Re$(\sigma)~ [\Lambda^2 \mu]$ & -Im$(\sigma) ~[\Lambda^2 \mu]$  \\
\hline
 0.25636863 & 0.05347248  \\
 0.32633833  &0.10252773 \\
0.47822011  & 0.10629266 \\
\hline
 0.26620715 & 0.02511717 \\
 0.32967925 & 0.08729944\\
0.41859618 & 	 0.08681749 \\
\hline
 \end{tabular}
\end{table*}

Note that this class of BS modes is qualitatively similar to the $w$ modes of constant density stars with comparable compactness~\cite{Kokkotas:1999bd}.
Computing the modes for the mini BS and massive BS models is more challenging because the imaginary part of these modes is comparable to the real part. In this case, a direct integration method becomes inaccurate. On the other hand, for these cases, we have successfully implemented the continued fraction method discussed above. Some modes for the mini BS model and the massive BS model are presented in Table~\ref{tab:BS_axial}. We note that, according to Ref. \cite{Benhar:1998au}, the value of $R_2$ in the expansion~\eqref{subst} cannot be completely arbitrary. In fact, it has to be slightly larger than the BS effective radius, in order to obtain a stable mode. This introduces an intrinsic inaccuracy in the BS QNMs computed with the continued fractions. Indeed, at $r=R_2$ the background scalar field is not exactly vanishing and the recursion relations~\eqref{4trrgiusta} are not exactly satisfied. This error decreases for compact configurations because the scalar field decays faster. In our calculations, we use $R_2=1.4 R$ and check the accuracy of the method by changing the location of $R_2$ in the range $1.3M$ to $1.5M$. We estimate an error of a few percent in the values presented in Table~\ref{tab:BS_axial}.
\begin{table*}
\caption{Axial QNMs of mini BS and massive BS configurations for $l=1$ and $l=2$, computed by a continued fraction method.}\label{tab:BS_axial}
$l=1$\hspace{6cm}$l=2$\\
 \begin{tabular}{| c | c | c | c | c | c |}
 \hline
 Model & N & Re$(\sigma)~ [\mu]$ & -Im$(\sigma) ~[\mu]$ & Re$(M\sigma)$ & -Im$(M\sigma)$ \\
 \hline
Mini BS I & 1 & 0.136 & 0.254 & 0.085  & 0.160 \\
Mini BS I & 2 & 0.316 & 0.388 & 0.200  & 0.245 \\
\hline
Mini BS II & 1 & 0.297 & 0.296 & 0.158  & 0.158 \\
Mini BS II & 2 & 0.725 & 0.457 & 0.387 & 0.244 \\
\hline 
Massive BS I & 1 & 0.228 & 0.207 & 0.515  & 0.468 \\
Massive BS I & 2 & 0.416 & 0.184 & 0.940  & 0.415 \\
\hline
Massive BS II & 1 & 0.264 & 0.213 & 0.508  & 0.410 \\
Massive BS II & 2 & 0.473 & 0.190 & 0.913 & 0.366 \\
\hline 
 \end{tabular}
 \begin{tabular}{| c | c | c | c | c | c |}
 \hline
 Re$(\sigma)~ [\mu]$ & -Im$(\sigma) ~[\mu]$ & Re$(M\sigma)$ & -Im$(M\sigma)$ \\
 \hline
 0.277 & 0.388 & 0.175  & 0.246\\
 0.456 & 0.374 & 0.289  & 0.237\\
\hline
 0.452 & 0.552 & 0.242  & 0.295\\
 0.721 & 0.456 & 0.385 & 0.244\\
\hline 
 0.225 & 0.197 &0.507  & 0.444\\
 0.375 & 0.180 & 0.847  & 0.408\\
\hline
 0.260 & 0.204 & 0.502  & 0.395\\
 0.437 & 0.182 & 0.844 & 0.351\\
\hline 
 \end{tabular}
\end{table*}

Let us now discuss the polar modes, which show a much richer structure due to the coupling between gravitational and scalar perturbations. Some of these modes were computed in Ref.~\cite{Yoshida:1994xi} using a WKB approximation, for the cases in which $\mu<(\sigma\pm\omega)$ [cf. Eq.~\eqref{kpm}]. In this case, both gravitational and scalar perturbations behave as outgoing waves at infinity. However, this restriction prevents the existence of quasibound-state modes for the scalar field perturbations, which are expected to dominate in the late time signal. Here we focus on this complementary regime, where scalar perturbations admit localized states (cf. Table~\ref{table:BCs}). We have obtained the fundamental modes of our BS models using the direct integration method described above. A selection of the results is presented in Table~\ref{tab:BS_polar}. For the solitonic BS polar modes, due to the precision needed for the background, a precise root finder method was not possible, making the modes more inaccurate than the mini and massive BS cases.

\begin{table*}
\caption{Polar QNMs of mini BS, massive BS and solitonic BS configurations for $l=0$ (left), and $l=2$ (right), computed by a direct integration method.}\label{tab:BS_polar}
$l=0$\hspace{6cm}$l=2$\\
 \begin{tabular}{| c | c | c | c | c | c |}
\hline
 Model & N & Re$(\sigma)~ [\mu]$ & -Im$(\sigma) ~[\mu]$ & Re$(M\sigma)$ & -Im$(M\sigma)$ \\
\hline
Mini BS I & 1 & 0.001416 & $1\times 10^{-11}$ & 0.0009  & $7\times 10^{-12}$ \\
Mini BS I & 2 & 0.11356 & $1\times 10^{-13}$ & 0.0719  &  $9\times 10^{-14}$\\
Mini BS I & 3 & 0.12958 & $9\times 10^{-15}$ & 0.0820  &  $5\times 10^{-15}$\\
\hline 
Massive BS I & 1 & 0.0197 & $1\times 10^{-4}$ & 0.04460  & $4\times 10^{-4}$\\
Massive BS I & 2 & 0.0636 & $ 1\times 10^{-11}$ & 0.1436  & $4\times 10^{-11}$\\
Massive BS I & 3 & 0.0896 & $5\times 10^{-13}$ & 0.2023  & $4\times 10^{-13}$\\
\hline 
Solitonic BS I & 1 &$ 3\times 10^{-4}$ & $2\times 10^{-5}$  & 0.002  & $1\times 10^{-4}$\\
Solitonic BS I & 2 & 0.063           & $9\times 10^{-6}$  & 4.631  & $7\times 10^{-5}$\\
Solitonic BS I & 3 & 0.103           & $2\times 10^{-10}$ & 7.601  & $2\times 10^{-9}$\\
\hline 
 \end{tabular}
 \begin{tabular}{| c | c | c | c | }
\hline
Re$(\sigma)~ [\mu]$ & -Im$(\sigma) ~[\mu]$ & Re$(M\sigma)$ & -Im$(M\sigma)$ \\\hline
 0.1195 & $5\times 10^{-5}$ & 0.0757  & $3\times 10^{-5}$ \\
 0.1316 & $2\times 10^{-5}$ & 0.0833  &  $1\times 10^{-5}$\\
0.1404 & $8\times 10^{-6}$ & 0.0888  &  $5\times 10^{-6}$\\
\hline 
0.0403 & $2\times 10^{-5}$ & 0.0909  & $6\times 10^{-5}$\\
 0.0716 & $2\times 10^{-6}$ & 0.1616  & $5\times 10^{-6}$\\
0.0947 & $5\times 10^{-7}$ & 0.2136  & $1\times 10^{-7}$\\
\hline 
 0.0348 & $1\times 10^{-4}$ & 0.3137  & $1\times 10^{-3}$\\
 0.0769 & $3\times 10^{-5}$ & 0.6928  & $2\times 10^{-4}$\\
 0.1127 & $4\times 10^{-6}$ & 1.0156  & $3\times 10^{-5}$\\
\hline
 
 \end{tabular}

\end{table*}

In Tables~\ref{tab:BS_axial} and \ref{tab:BS_polar} we also show the $l=1$ axial modes and the $l=0$ polar modes, respectively. Given the quadrupolar nature of GR, in the Schwarzschild case the $l=0,1$ perturbations are simply associated with infinitesimal changes in the mass and in the angular momentum, respectively~\cite{Regge:1957td,Zerilli:1971wd}. However, due to the coupling with the scalar field, for BSs these modes become part of the spectrum and are associated with monopole and dipole emission.

Finally, by comparing the real part of the polar modes shown in Table~\ref{tab:BS_polar} with the orbital frequency of circular geodesics shown in Fig.~\ref{fig:circular_geodesics}, we observe that such modes can be potentially excited by a quasicircular EMRI~\cite{Pons:2001xs,Pani:2010em} in the point-particle limit. We investigate this effect in the next section.

\section{Point particle orbiting a boson star \label{sec:PP}}
The gravitational and the scalar wave emission by a particle in a circular geodesic motion around a BS is governed by the inhomogeneous system of equations~\eqref{rweq} and~\eqref{polareq1}-\eqref{polareq3} and \eqref{eqphiper}. The solutions can be constructed via Green's function techniques. Once again, we shall treat the axial and polar sectors separately. 
\subsection{Axial sector}
\begin{figure}
\begin{center}
\begin{tabular}{c}
\epsfig{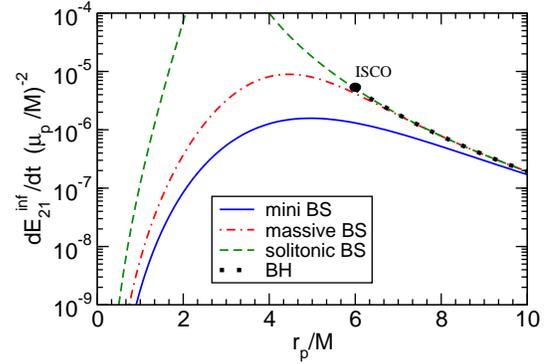}
\end{tabular}
\end{center}
\caption{\label{fig:axial_flux}
(Adapted from Ref.~\cite{Macedo:2013qea})
Dominant $l=2$, $m=1$ contribution to the axial gravitational flux emitted by a point particle orbiting a BS for the stable BS configurations used in this work, compared to that of a Schwarzschild BH. The solitonic configurations for $r>3M$ have basically the same values as the BH case.
}
\end{figure}

The axial sector is fully described by Eq.~\eqref{rweq}. The general solution can be constructed from two independent solutions of the associated homogeneous equations:
\bea\label{axialgen}
\Psi_{RW}=\frac{1}{W_{Z}}\l[Z_+(r)\int_{0}^r dr_*Z_- S_{RW} +\r.\nn\\\l.Z_-(r)\int_{r}^\infty dr_*Z_+ S_{RW} \r],
\eea
where $Z_{\pm}$ are solutions of the homogeneous associated equation with the following boundary conditions
\bea
Z_{+}(r\rightarrow \infty)&\sim& e^{i\sigma r_*},\\
Z_{-}(r\rightarrow 0)&\sim& r^{l+1},
\eea
and $W_Z=Z_-(dZ_+/dr_*)-Z_+(dZ_-/dr_*)$ is the Wronskian. At large distance, the solution~\eqref{axialgen} reads
\be
\Psi_{RW}(r\rightarrow \infty)\sim\frac{e^{i\sigma r_*}}{W_{Z}}\int_{0}^\infty dr_*Z_- S_{RW}.
\label{rwinf}
\ee
For circular orbits the source terms generically contain Dirac's delta terms $\delta(r-r_p)$ and their derivative, namely:
\be
S_{RW}= \left[G_{RW}\delta(r-r_p)+F_{RW} \delta'(r-r_p)\right]\delta(\sigma-m\Omega)\,,\nn
\ee
so that the solution~\eqref{rwinf} can be rewritten as
\be
\Psi_{RW}\sim \bar{\Psi}_{RW} \delta(\sigma-m\Omega) e^{i \sigma r_*}\,,
\ee
where~\cite{Martel:2003jj}
\bea
\bar{\Psi}_{RW}=\l.\frac{e^{\frac{1}{2}(u-v)}}{W_{Z}}\l[ G_{RW}Z_- -\frac{d}{dr_*}\l(e^{\frac{1}{2}(u-v)} F_{RW}Z_-\r)\r]\r|_{r=r_p}\,.\nn
\eea
Finally, the energy flux at (null) infinity due to the axial part of the perturbations is given by \cite{Martel:2003jj,Martel:2005ir}
\be
\dot{E}^{\text{inf,axial}}_{lm}=\frac{1}{16 \pi}\frac{(l+2)!}{(l-2)!}\l|\bar{\Psi}_{RW}\r|^2. \nn
\ee
Due to the explicit form of the source term, the axial flux is vanishing for even values of $l+m$.
In Fig.~\ref{fig:axial_flux} we show the dominant $l=2$, $m=1$ contribution of the axial flux for various stable BS models as well as that of a Schwarzschild BH. The deviations from the BH case are basically indistinguishable at large distances. As expected, more compact configurations like the solitonic BS model are closer to the BH case.
\subsection{Polar sector}
The polar sector is described by the inhomogeneous system of coupled equations~\eqref{polareq1}--\eqref{polareq3}. A general method to solve this class of problems was presented in Ref.~\cite{Pani:2011xj}, which we shall closely follow. The polar equations can be written as
\be
\frac{d\mathbf{\Psi}}{dr}+\mathbf{V} \mathbf{\Psi}=\mathbf{S},
\label{polarsystem}
\ee
where we introduced the six-dimensional vectors
\be
\mathbf{\Psi}=\l(H_1,K,\phi_+,\phi_-,\phi_+',\phi_-'\r),
\ee
and the vector $\mathbf{S}$ describes the source terms. The matrix $\mathbf{V}$ can be straightforwardly constructed from Eqs.~\eqref{polareq1}--\eqref{polareq3}. In order to solve Eq.~\eqref{polarsystem}, let us define the $6\times6$ matrix $\mathbf{X}$, whose columns are formed by independent solutions of the associated homogeneous problem. It is easy to show that
\be
\frac{d\mathbf{X}}{dr}+\mathbf{V} \mathbf{X}=0.
\ee
The general solution can be written in terms of the homogeneous solutions by~\cite{Pani:2011xj}
\be
\mathbf{\Psi}=\mathbf{X}\int dr\mathbf{X}^{-1}\mathbf{S}.
\ee
The matrix $\mathbf{X}$ can be constructed in the following way~\cite{Molina:2010fb,Pani:2011xj}: the solution close to the origin is defined by three independent parameters, say $(\psi_0^{or},\phi_+^{or},\phi_-^{or})$. Likewise, the solution close to infinity is characterized by $(\psi_0^{\infty},\phi_+^{\infty},\phi_-^{\infty})$. We can construct three independent solutions integrating the equations from the origin by setting the triad to $(1,0,0)$, $(0,1,0)$ and $(0,0,1)$. Using the same for the integration from infinity, we construct the set of six independent solutions which form $\mathbf{X}$.

The boundary conditions for the problem are analogous to those described in the previous sections. For the gravitational functions, we require regularity at the origin and outgoing waves at infinity. For the scalar field, we require regularity at the origin, but the condition of outgoing waves is not satisfied for all values of $\Omega$. In fact, for sufficiently small frequencies, when $k_\pm^2 >0$ [cf. Eq.~\eqref{kpm}], the perturbations of the scalar field are localized near the star and form quasibound states. If $k_\pm^2 <0$, the orbital frequency is larger than the potential well and the perturbations are wavelike at infinity. 
The value of $\Omega$ for which this transition occurs depends on the specific model through $\mu$ and $\omega$ and on the azimuthal number $m$ [cf. Eq.~\eqref{kpm} and recall that, for a circular orbit, $\sigma = m \Omega$].

To compute the polar gravitational part of the flux, we construct the Zerilli function at infinity, using the solutions for $K$ and $H_1$ obtained by solving the coupled system. Then, the polar gravitational flux is the sum of the multipolar contributions \cite{Martel:2003jj,Martel:2005ir,Pani:2011xj}:
\be
\dot{E}^{\text{inf,}Z}_{lm}=\frac{1}{64 \pi}\frac{(l+2)!}{(l-2)!}(m\Omega_p)^2\l|\Psi_{Z}(r\rightarrow \infty)\r|^2\,,
\ee
which, by virtue of the specific source term, are nonvanishing only for even values of $l+m$.

The scalar flux can be computed through the energy momentum tensor of the scalar field~\cite{Breuer:1974uc,Pani:2011xj} (see also~Refs. \cite{Crispino:2000am,Crispino:2008zza} for another approach). It reads
\be
\dot{E}^{\text{inf},\phi_\pm}_{lm}=2 (m\Omega_p)^2|{\phi}_\pm(r\rightarrow \infty)|^2.
\ee
The total energy flux for the polar sector is the sum of the two contributions, i.e.
\be
\dot{E}^{\text{inf,polar}}_{lm}=\dot{E}^{\text{inf,}Z}_{lm}+\dot{E}^{\text{inf},\phi_\pm}_{lm} \,.
\ee
In the next subsection, we give the details of the polar part of the flux.

\subsection{Emitted polar flux and inspiral resonances}

Adopting the procedure explained above, we have evaluated the total scalar and polar gravitational flux emitted by a test particle orbiting a BS in several BS models.
In some cases, the numerical integration is challenging. Indeed, for sufficiently small orbital frequency the scalar perturbations decay exponentially at infinity, but they are nonetheless coupled to the gravitational perturbation which instead propagate to infinity as waves. To achieve good accuracy, the numerical domain of integration should extend up to many wavelengths, i.e. $r_\infty\sigma\gg1$, where $r_\infty$ is our numerical value for infinity. On the other hand, the typical length scale of the scalar perturbation is given by the Yukawa-like term, i.e. $1/\mu$. Due to the exponential decay, it is challenging to integrate the scalar field if $r_\infty\mu\gg1$, and this sets a limit to the values of $r_\infty$ that can be used. To circumvent this problem, we have constructed the large distance solution perturbatively using many terms (typically $20$) in the series expansion of the solutions at the infinity. This allows to reduce numerical truncation errors. Note that this problem becomes more severe when the mass of the scalar field is large, $\mu\gg\sigma$.

\begin{figure*}
\begin{center}
\epsfig{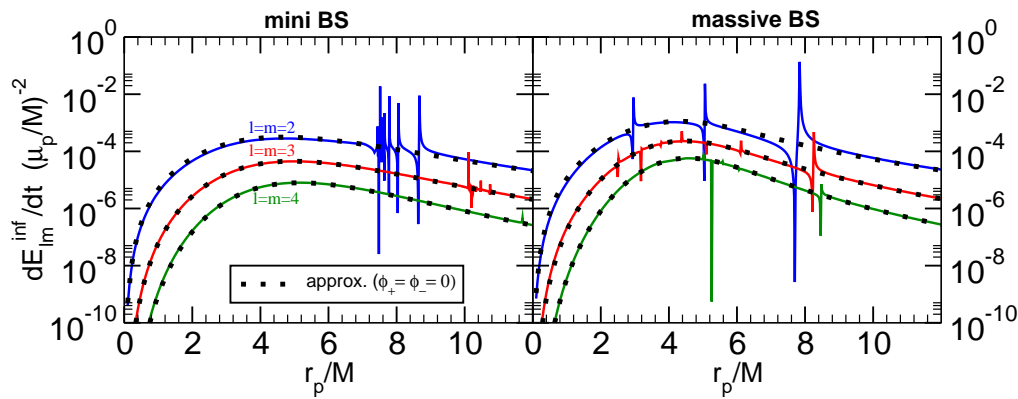}
\end{center}
\caption{\label{fig:flux_polar}
(Adapted from Ref.~\cite{Macedo:2013qea})
Main multipole contributions to the polar flux,  $l,m=2,3$ and $4$, for the mini and massive BS configurations I. The dots indicate the approximated results obtained by setting the scalar perturbations to zero.}
\end{figure*}
An interesting phenomenon that occurs for test particles orbiting a relativistic star is the appearance of resonances in the flux (see, e.g., Ref.~\cite{Pons:2001xs}). The resonance condition reads
\be
m \Omega =\sigma_R,
\ee
where $m$ is the azimuthal number and $\sigma_R$ is the real part of the QNM frequency. In other words, if the characteristic frequency of the BS matches (multiples of) the orbital frequency of the particle, sharp peaks appear in the emitted flux. This is consistent with a simple harmonic oscillator model, where the orbiting particle acts as an external force and where $\sigma_R$ is the proper frequency of the system. In this picture, the imaginary part of the frequency $\sigma_I$ is related to the damping of the oscillator and it is roughly proportional to the width of the resonance, while the quality factor $\sigma_R/\sigma_I$ is proportional to the square root of the height~\cite{Pons:2001xs}.

The appearance of these resonances seems to be a generic feature of BSs.
As shown in Fig.~\ref{fig:flux_polar}, the resonant frequencies may correspond to a stable circular orbit located \emph{outside} the BS effective radius (as for the rightmost resonance of the massive BS case in the right panel of Fig.~\ref{fig:flux_polar}) or may correspond to stable circular orbits \emph{inside} the BS (as in the mini BS case shown in the left panel of Fig.~\ref{fig:flux_polar}). While resonant circular orbits also occur outside perfect-fluid stars~\cite{Pons:2001xs} and gravastars~\cite{Pani:2010em}, the existence of resonant geodesics \emph{inside} the compact object is peculiar of BSs, due to the absence of a well-defined surface and due to the existence of \emph{stable} circular orbits inside the star~\cite{Macedo:2013qea}. We shall address the solitonic BS case later, due to its complexity.

The existence of these inner resonances is intriguing because they appear to be a generic feature of compact objects supported solely by the self-gravity of a scalar field. Indeed, any sufficiently compact object can support bound and quasibound modes in its interior. In Appendix~\ref{app:stars}, we show that constant density stars can support bound-state modes (i.e. modes with purely real frequency) for massive scalar perturbations with $l>0$, and they can also support quasibound modes (i.e. modes with small but nonvanishing imaginary part) for massless scalar and for gravitational perturbations. In the case of ordinary stars, these modes cannot be excited because their frequency is higher than the frequency of the innermost stable circular orbit. However, the same class of modes exists also for BSs which, however, admit stable circular orbits in their interior. In the case of a BS, even the massive scalar modes are quasibound. The small imaginary part of the frequency is related to the coupling between scalar and gravitational perturbations: even if the scalar flux is zero for bound-state modes, part of the energy carried by the scalar field can be converted into gravitational energy that is then dissipated at infinity through gravitational waves. This also explains qualitatively why the imaginary part of these modes is small (i.e. why the resonances are generically narrow) because the dissipation mechanism is not efficient. 

\begin{figure*}
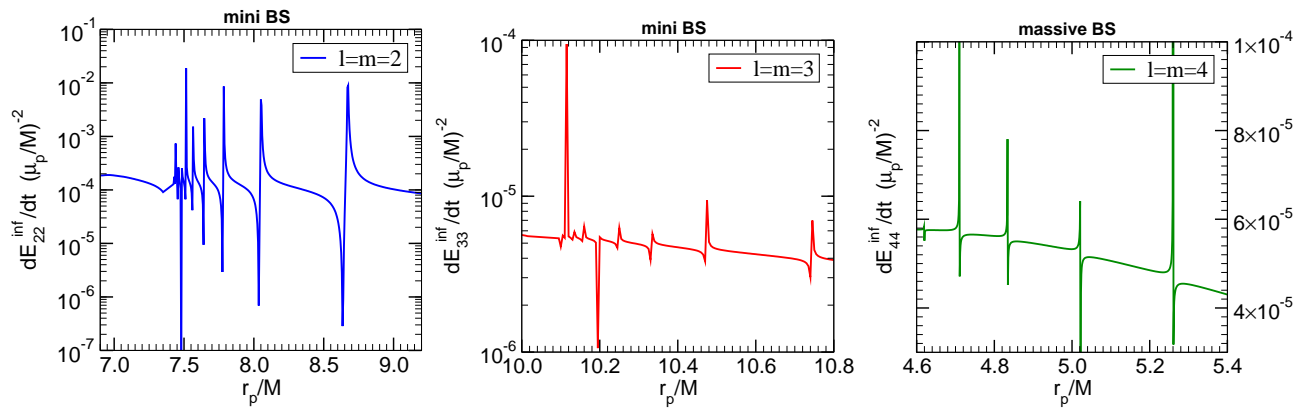

 \begin{center}
 \begin{tabular}{ccc}
 \epsfig{file=Plots/polarflux_zoom1.eps,width=5.5cm,angle=0,clip=true}& 
 \epsfig{file=Plots/polarflux_zoom2.eps,width=5.6cm,angle=0,clip=true}&
  \epsfig{file=Plots/polarflux_zoom3.eps,width=5.7cm,angle=0,clip=true}
 \end{tabular}
 \end{center}
\caption{\label{fig:flux_polar_zoom}
 Zoom of the main multipole contributions to the polar flux for orbital frequencies close to the interface condition~\eqref{Omegares}. Left panel: mini BS configuration~I for $l=m=2$. In this case the interface condition corresponds to $r_p=7.3624M$. Middle panel: Mini BS configuration~I for $l=m=3$; the interface condition corresponds to $r_p=10.0292 M$. Right panel: Massive BS configuration~I for $l=m=4$; the interface condition corresponds to $r_p=3.8540 M$.}
\end{figure*}

The structure of the resonances is fairly rich and it depends on the values of $l$, $m$ and on the specific BS model. We can gain some insight by looking at the analog problem for a Schwarzschild BH. In that case, the location and width of the resonances can be computed analytically in the small mass limit~\cite{Detweiler:1980uk,Cardoso:2011xi}. For the Schwarzschild BH case, the real and imaginary part of the quasi-bound modes read
\begin{eqnarray}
 \sigma_R&\approx&\mu\left(1-\frac{M^2\mu^2}{2 (n+l+1)}\right)\,,\nn\\
 \sigma_I&\approx&-\frac{4^{1-2 l}\pi ^2 (M\mu)^{4l+6}}{M(1+l+n)^{2 (2+l)}}\left[\frac{ (2l+n+1)!}{ \Gamma\left[\frac{1}{2}+l\right]^2 \Gamma\left[\frac{3}{2}+l\right]^2 n!}\right]\,,\nn
\end{eqnarray}
where $n\geq0$ is the overtone number. Therefore, as $\sigma$ approaches $\sigma_R$ there is a multitude of modes that can be excited and their separation in orbital frequency vanishes in the large $l$ or large $n$ limit. However, in the same limit the imaginary part (and hence the width of the resonances) of the modes decreases very rapidly, as shown by the last equation above. Our results for the resonances appearing in the flux from a BS inspiral are in qualitative agreement with this behavior.  This is shown in Fig.~\ref{fig:flux_polar_zoom}, where we show the polar flux in a restricted region of the orbital radius for some BS model. 
Due to the complex scalar field, the resonance condition is shifted, $\sigma\pm\omega\approx\mu$, and corresponds to $k_\pm\approx0$ in Eq.~\eqref{kpm}, i.e. to the interface between quasibound states and QNMs.
In the left panel of Fig.~\ref{fig:flux_polar_zoom}, we show the main $l=m=2$ contribution for our mini BS configuration~I. In this case, the interface condition $k_+=0$ occurs at $r_p\approx 7.3624 M$ and, even for $l=2$, several resonances appear when the particle approaches this peculiar orbit. Similar results hold for the contribution to the flux $l=m=3$ for the mini BS configuration I and $l=m=4$ for the massive BS configuration~I. In these cases, the interface conditions read $r_p\approx 10.0292 M$ and $r_p\approx 3.8540 M$, respectively. Note that the width of the resonances decreases very rapidly for large values of $l$, so that the resonances of higher multipoles are more difficult to resolve and the corresponding modes have a smaller quality factor.
%

Our analysis generically shows that the orbital frequency
\begin{equation}
 \Omega_{\rm res}\sim\frac{\mu\mp\omega}{m}\,,\label{Omegares}
\end{equation}
plays a special role in the gravitational and scalar flux emitted in a quasicircular inspiral around a BS. The detectability and some observational implications of these resonant frequencies are discussed in Ref.~\cite{Macedo:2013qea}.

\begin{figure}
\begin{center}
\epsfig{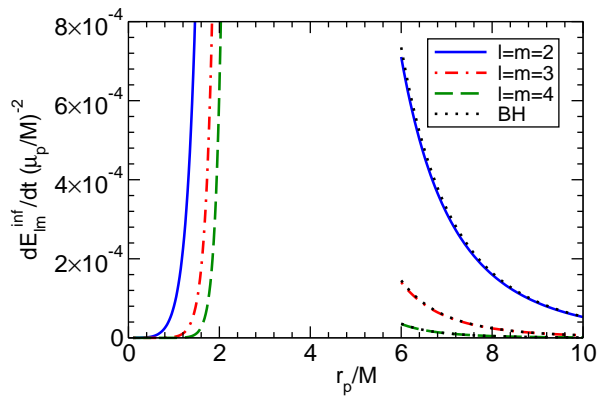}
\end{center}
\caption{\label{fig:flux_polar_solitonic}
Polar flux for the stable {(solitonic BS I configuration, cf. Table \ref{tableconf})}, compared with the Schwarzschild BH case.
}
\end{figure}
The presence of the scalar field perturbations is crucial for the resonances. In order to illustrate this point, we have considered a decoupling limit, where gravitational and scalar perturbations do not couple to each other.  Although this approximation is not fully consistent, it is nevertheless useful to separate the features of the flux computed for the full coupled system. In this limit, scalar perturbations are described by two coupled second-order equations, which support \emph{normal} modes, i.e. modes with a purely real part. These modes are very close to the real part of the slowly damped modes found in the full system and they roughly coincide with the resonant frequencies of the point-particle quasicircular inspiral. 
Likewise, we have computed the gravitational flux after having artificially set the scalar perturbations to zero. In Fig.~\ref{fig:flux_polar} we show a comparison between the fluxes in the decoupling limit and those obtained by solving the full equations for mini BSs and massive BSs. Away from the resonances, the gravitational flux is in very good agreement with the exact result. This is consistent with the picture presented above: in the absence of gravito-scalar coupling, the system would admit normal bound scalar modes. The later, however, acquire a small imaginary part due to the coupling with the gravitational sector and can be dissipated at infinity as gravitational waves. Thus, the real part of the QNMs is mainly governed by the scalar sector, whereas the generic aspects of the flux away from the resonant modes is mainly driven by the gravitational sector.

Supported by the good agreement of the decoupling of the scalar field and gravitational perturbations, we have adopted it to compute the flux in the solitonic BS configurations. In this case, the large mass of the background scalar field makes it challenging to solve the full system. This is due to the presence of two different length scales: the BS mass $M$ which regulates the gravitational sector, and the scalar field mass $\mu M\gg1$ which regulates the decay of the scalar field.

In Fig.~\ref{fig:flux_polar_solitonic} we show the flux obtained in the decoupling limit, compared to its (exact) Schwarzschild counterpart. {We only show the stable circular orbits located roughly at $r>6M$}. For these orbits, the difference is small. This is expected because the external spacetime is very close to the Schwarzschild one. On the other hand, unlike the other BS models, highly energetic stable circular orbits exist close to the stable light ring inside the star. In Fig.~\ref{fig:flux_polar_solitonic} we have neglected the resonance structure of the flux. However, as shown in Table~\ref{tab:BS_polar}, resonant frequencies for this model correspond to  relativistic high-energy orbits and they are not excited by the quasicircular inspiral. Furthermore, since the mass coupling of these configurations is higher than the other models, scalar field radiation is only emitted for very high multipoles, which are subdominant. Thus, for solitonic compact configurations the main distinctive feature of such compact horizonless configurations with respect to a Schwarzschild BH is the possibility of having stable geodesics in the core of the object, which are also associated with large gravitational fluxes. We refer the reader to Ref.~\cite{Macedo:2013qea}, where other features of the inner inspiral are discussed.
\section{Conclusions and outlook}

In this work we constructed three different BS models, namely, mini, massive and solitonic BSs. 
The spacetimes were constructed using the full Einstein equations, without any approximation method. 
Moreover, we discussed circular geodesic motion in the BS spacetime, showing some specific features that are also present in the case of circular motion in uniform density stars in general relativity, like the presence of two light rings, depending on the compactness of the star. We computed the QNMs of the BS configurations, extending the results of Ref.~\cite{Yoshida:1994xi}, showing that generically they would be excited by the motion of a point particle in circular orbits. The energy fluxes emitted by the particle were calculated, showing the distinctive characteristics of the resonances in the flux. The analysis made here also extends the results of Ref.~\cite{Kesden:2004qx}. The discussion on the detectability and observational consequences of the resonances was given in Ref.~\cite{Macedo:2013qea}.

The results presented in this paper offer an answer to the question of whether or not one can distinguish BSs from BHs, from the gravitational point of view. We conclude that the motion of stellar-size objects would leave characteristic imprints in the signal that are intrinsically connected with the BS models. The mass of the bosonic particle forming the star has to be light enough  to reproduce supermassive objects. The studies presented here for the gravitational flux are for point particles in circular orbits, and are most applicable in the region where the scalar field $\phi_0$ is small enough, i.e., outside an effective BS radius. Inside the star other effects like accretion and dynamical friction should be considered (see \cite{Macedo:2013qea} for more details). In particular, these effects in a head-on collision could lead to interesting features.
Also, the study of eccentric orbits is a direct and important generalization of the present work.

Another possible extension of the present study is the investigation of EMRIs in rotating BS spacetimes. Rotating BSs were analyzed in the literature, in both Newton's and Einstein's gravity context \cite{Yoshida:1997qf}. Perturbation theory around nonspherically symmetric spacetimes is still a challenge, and some cases were studied only within approximation schemes, like in slowly rotating BHs \cite{Pani:2012bp,Pani:2013ija,Pani:2013pma}. The study presented here serves as a reference for further studies of BSs systems. 

\begin{acknowledgments}
We are indebted to Emanuele Berti, Kazunari Eda, Luis Lehner, Joseph Silk and especially to Enrico Barausse for useful correspondence.
C. M. and L. C. acknowledge CAPES and CNPq for partial financial support. L. C. is grateful to CENTRA-IST for kind hospitality.
P.P. acknowledges financial support provided by the European Community 
through the Intra-European Marie Curie Contract No. aStronGR-2011-298297 
and the kind hospitality of Kinki University in Osaka.
V.C. acknowledges partial financial
support provided under the European Union's FP7 ERC Starting Grant ``The dynamics of black holes:
Testing the limits of Einstein's theory'' Grant Agreement No. DyBHo--256667.
Research at Perimeter Institute is supported by the Government of Canada 
through Industry Canada and by the Province of Ontario through the Ministry
of Economic Development and Innovation.
This work was supported by the NRHEP 295189 FP7-PEOPLE-2011-IRSES Grant, and by FCT-Portugal through Projects No.
PTDC/FIS/098025/2008, No. PTDC/FIS/098032/2008, No. PTDC/FIS/116625/2010, No.
CERN/FP/116341/2010 and No. CERN/FP/123593/2011.
Computations were performed on the ``Baltasar Sete-Sois'' cluster at IST,
the cane cluster in Poland through PRACE DECI-7 ``Black hole dynamics
in metric theories of gravity,''  
on Altamira in Cantabria through BSC Grant No. AECT-2012-3-0012,
on Caesaraugusta in Zaragoza through BSC Grants No. AECT-2012-2-0014 and No. AECT-2012-3-0011,
XSEDE clusters SDSC Trestles and NICS Kraken
through NSF Grant~No.~PHY-090003, and Finis Terrae through Grant No.
CESGA-ICTS-234.
\end{acknowledgments}

\appendix

\section{Massive scalar modes of a constant density star}\label{app:stars}
In this appendix we compute the massive scalar modes of a constant density star and show that they share many features with those obtained for the BS models presented in the main text.
The background metric of a spherically symmetric star reads
\begin{equation}
 ds_0^2=-e^{v(r)}dt^2+e^{u(r)}dr^2+r^2(d\theta^2+\sin^2\theta d\varphi^2)\,,
\end{equation}
where $e^{-u(r)}=1-2m(r)/r$. In the case of an isotropic, perfect-fluid star, the Einstein's equations are given by~\cite{Shapiro:1983du}:
\bea
m'(r)&=&4 \pi r^2 \rho(r),\label{eq:fluid1}\\
v'(r)&=&2\frac{m(r)+4 \pi  r^3 P(r)}{r^2-2 r m(r)},\label{eq:fluid2}\\
P'(r)&=&-\frac{\left(m(r)+4 \pi  r^3 P(r)\right) (P(r)+\rho (r))}{r (r-2 m(r))}\label{eq:fluid3},
\eea
together with an equation of state, relating $\rho$ with $P$.

We consider a probe scalar field which satisfies the massive Klein-Gordon equation $\square\psi-\mu^2\psi=0$. In order to facilitate a comparison with the BS cases, we assume an ansatz $\psi=\Psi(r) r^{-1} Y_{lm} e^{i(\sigma\pm\omega)t}$. The scalar perturbation equation then reads
\begin{equation}
\frac{d^2}{dx^2}\Psi+\left[(\sigma\pm\omega)^2-V_0\right]\Psi=0\,,
\end{equation}
with
\begin{equation}
 V_0=e^{v}\left(\mu ^2+\frac{l (l+1)}{r^2}+\frac{2 m}{r^3}+4 \pi ( P-\rho)\right)\,,\label{eq:scalarpot}
\end{equation}
where we have used Eqs.~\eqref{eq:fluid1} and \eqref{eq:fluid2} in order to eliminate $v'$ and $m'$.

For constant density stars we have that $\rho(r)=\rho_c$, and Eqs.~\eqref{eq:fluid1}--\eqref{eq:fluid3} can be solved analytically, resulting in
\bea
m&=& \frac{4}{3}\pi r^3 \rho_c,\\
e^{v}&=&\left[\frac{3}{2}\left(1-\frac{2M}{R}\right)^{1/2}-\frac{1}{2}\left(1-\frac{2Mr^2}{R^3}\right)^{1/2}\right]^2,\\
P&=&\rho_c\l[\frac{R(R-2M)^{1/2}-(R^3-2Mr^2)^{1/2}}{(R^3-2Mr^2)^{1/2}-3R(R-2M)^{1/2}}\r].
\eea
In the equations above, $R$ is the star radius and $M=m(R)$ is the total mass. The solution above is valid for $r<R$, whereas for $r>R$ the spacetime coincides with the Schwarzschild one due to Birkhoff's theorem.

\begin{figure}
\begin{center}
\epsfig{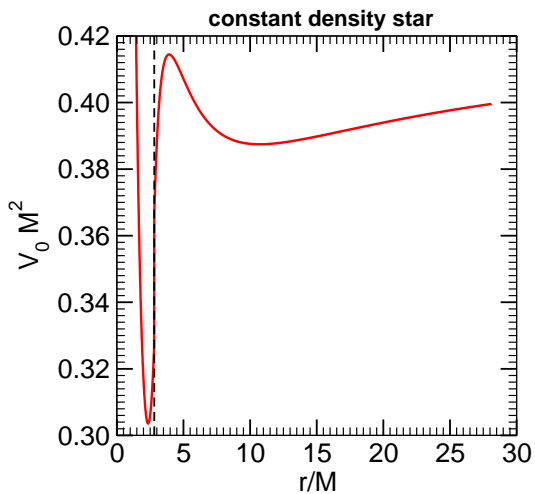}
\end{center}
\caption{\label{fig:potential_star}
Example of Schroedinger-like potential for a massive scalar perturbation of a constant density star with $l=2$, $R/M\approx2.81$ and $M\mu=0.65$. For this choice of parameters, the potential has two minima. The location of the star radius is marked by the vertical dashed line.
}
\end{figure}

For constant density stars, the potential \eqref{eq:scalarpot} can support bound states in a certain region of the $\mu$--$M$ parameter space. An example is shown in Fig.~\ref{fig:potential_star}.

The potential may develop up to two minima: one is located in the outer region for a certain range of nonvanishing $\mu$ and exists for sufficiently compact stars; the other is located inside the star and it also exists also at small densities if the scalar mass $\mu$ is sufficiently large. Furthermore, the inner minimum also exists when $\mu=0$ in a small range of compactness. In both cases, the system allows for \emph{normal}, bound modes, i.e. modes characterized by a purely real frequency, which can be straightforwardly computed. In Table~\ref{tab:stars} we show some modes computed using a direct integration method for $l=2$, $R/M\approx 6.93$, $M\mu\approx 2.257$ and $M\omega=1.865$ and $l=2$, $R/M\approx 3.10$, $M\mu\approx 7.37$ and $M\omega=0.899$.
\begin{table}
\caption{A selection of massive scalar modes $\omega_n$ of a constant density star for $l=2$. The parameters chosen in the left part of the table are $R/M\approx 6.93$, $M\mu\approx 2.257$ and $M\omega=1.865$. For the right part of the table we have chosen $R/M\approx 3.10$, $M\mu\approx 7.37$ and $M\omega=0.899$. We adopt this choice of parameters to represent the massive BS I and solitonic BS I.}
\begin{tabular}{c | c | c  }
 \hline\hline
$n$ & $M\sigma_n$ & $M\Omega_p$ \\
\hline
1 & 0.030 & 0.015  \\
2 & 0.119 & 0.059  \\
3 & 0.188 & 0.094  \\
4 & 0.236 & 0.118  \\
5 & 0.271 & 0.135  \\
6 & 0.296 & 0.148  \\
7 & 0.314 & 0.157  \\
\hline\hline
\end{tabular}
\begin{tabular}{ c | c  }
 \hline\hline
 $M\sigma_n$ & $M\Omega_p$ \\
\hline
 2.632 & 1.316  \\
 2.850 & 1.425  \\
 3.065 & 1.532  \\
3.277& 1.638 \\
 3.486 & 1.743 \\
3.692 &1.846  \\
 3.891 & 1.945 \\
\hline\hline
\end{tabular}
\label{tab:stars}
\end{table}
These parameters were chosen to reproduce the massive and solitonic BS configuration I, analyzed in the main text (cf. Table~\ref{tableconf}). In those cases, the potential only has one minimum, located in the interior of the star. In Table \ref{tab:stars} we also exhibit the orbital frequency of a particle that excites the modes when $m=2$, i.e. when the condition $\sigma_n=2\Omega_p$ is met. This configuration is qualitatively similar to the case of a point-particle orbiting a BS, due to the coupling between scalar and gravitational perturbations. Indeed, the resonance frequencies are qualitatively similar to those obtained for the massive BS configuration~I in the main text. An important difference from the BS case is that even localized scalar modes acquire a small imaginary part. This is due to the fact that scalar perturbations are coupled to the gravitational ones and, although the former are localized in a region of width $\sim1/\mu$ close to the BS, the latter dissipate energy at infinity through gravitational-wave emission. Thus, part of the scalar field energy is converted and emitted as gravitational waves. We refer the reader to the discussion in the main text for the interpretation of these results and for more details.

\bibliography{emriboson}
\end{document}